\documentclass[manuscript]{aastex}

\slugcomment{Not to appear in Nonlearned J., 45.}

\shorttitle{The Starburst Open Cluster Westerlund 1}
\shortauthors{Lim et al.}

\begin{document}

\title{The Starburst Cluster Westerlund 1: \\
       The Initial Mass Function and Mass Segregation}

\author{Beomdu Lim\altaffilmark{1}, Moo-Young Chun\altaffilmark{2}, Hwankyung Sung\altaffilmark{1,5},
 Byeong-Gon Park\altaffilmark{2}, Jae-Joon Lee\altaffilmark{2},  
Sangmo T. Sohn\altaffilmark{3}, Hyeonoh Hur\altaffilmark{1}, and 
 Michael S. Bessell\altaffilmark{4}}
\email{bdlim1210@sju.ac.kr, sungh@sejong.ac.kr}

\altaffiltext{1}{Department of Astronomy and Space Science, Sejong University, 98 Gunja-dong, 
Gwangjin-gu, Seoul 143-747, Korea}
\altaffiltext{2}{Korea Astronomy and Space Science Institute, 776 
Daedeokdae-ro, Yusung-gu, Daejeon, Korea}
\altaffiltext{3}{Space Telescope Science Institute, Baltimore, MD 21218, USA}
\altaffiltext{4}{Research School of Astronomy and Astrophysics, Australian National University, 
MSO, Cotter Road, Weston, ACT 2611, Australia} 
\altaffiltext{5}{Corresponding Author}

\begin{abstract}
\objectname{Westerlund 1} is the most important starburst cluster in the 
Galaxy due to its massive star content. We have performed $BVI_C$ and $JK_{S}$ photometry 
to investigate the initial mass function (IMF). By comparing the observed color with 
the spectral type - intrinsic color relation, we obtain the mean interstellar reddening 
of $<E(B-V)>=4.19\pm0.23$ and $<E(J-K_{S})>=1.70\pm0.21$. Due to the heavy extinction toward the 
cluster, the zero-age main sequence fitting method based on optical photometry 
proved to be inappropriate for the distance determination, while the near-infrared photometry gave a reliable distance 
to the cluster -- 3.8 kpc from the empirical relation. Using the recent theoretical stellar
evolution models with rotation, the age of the cluster is estimated to be $5.0\pm1.0$ Myr. 
We derived the IMF in the massive part and obtained a fairly shallow slope 
of $\Gamma = -0.8\pm 0.1$. The integration of the IMF gave a total mass for the cluster in 
excess of $5.0 \times 10^4 M_{\sun}$. The IMF shows a clear radial variation indicating the 
presence of mass segregation. We also discuss the possible star formation history of Westerlund 1 
from the presence of red supergiants and relatively low-luminosity yellow hypergiants. 

\end{abstract}

\keywords{open clusters and associations: individual (Westerlund 1)--stars:luminosity function, mass function}

\section{INTRODUCTION}

Massive stars play a very significant role in the chemical evolution of a galaxy, as 
well as in the dynamical evolution of stellar systems. The formation mechanism of 
such massive stars still remains largely unknown, and should be explored in the future. 
The short lifetimes and rarity of massive stars make it 
difficult to study their evolution in detail. Currently the shape of the initial mass 
function (hereafter IMF) of low- to intermediate-mass stars is relatively well studied, while 
that of massive stars is poorly known due to their rarity. In this context, 
starburst clusters (or super star clusters) containing a few tens to hundreds of massive stars 
($\geq 15 M_{\sun}$), such as \objectname{Westerlund 1}, 
\objectname{the Arches}, \objectname{Quintuplet}, \objectname{NGC 3603}, and \objectname{Westerlund 2} in 
the Galaxy, and \objectname{R136} in the Large Magellanic Cloud, are ideal laboratories to study the issues mentioned 
above. Futhermore, systematic studies of such clusters may provide an opportunity to understand the 
star formation environment in starburst galaxies. Westerlund 1 is known as the most 
massive open cluster in the Galaxy \citep{CNCG05}, and is one of a few highly reddened 
clusters observable in the optical.

Westerlund 1 (also called the Ara cluster) is a young open cluster located in the 
constellation of Ara. \citet{W61} discovered the cluster with the 26 inch Uppsala-Schmidt telescope at 
Mount Stromlo Observatory. From multi-wavelength photographic observations he assigned the object as a 
heavily reddened cluster in the Sagittarius arm containing O, B, and M type stars with an age of 3 Myr. 
With 2.2 $\micron$ data from \citet{P68}, \citet{W68} also tested the extinction law. 
Since its discovery, several studies have only focused on the bright sources in the cluster 
using near-infrared (NIR) photometry. \citet{BKS70} observed 12 sources brighter than 5 mag in $K$ band and 
confirmed the presence of supergiants. \citet{LD74} also identified 
early and late type supergiants through observations with narrow-band NIR 
filters ($0.78-1.05 \micron$) and found that 
M type supergiants were more reddened than OB stars. The strong infrared excess 
of M type supergiants was also found by \citet{KF77} through photometry from 
the near to mid-infrared region ($2.2-20 \micron $). After a decade, 
\citet{W87} carried out photoelectric photometry and the first spectral classification of several 
bright stars in the cluster.  Toward the end of the 20th century, \citet{C98} studied two strong emission-line sources in the
radio and mid-infrared.  

In the 2000s, many significant discoveries were made from multi-wavelength studies. 
\citet{CN02, CN04}, motivated by the 
results of \citet{C98}, found a new luminous blue variable and eleven 
Wolf-Rayet stars from spectroscopic studies. Subsequently \citet{CNCG05} carried 
out photometric and spectroscopic observations with the 3.6m New Technology 
Telescope, and published new spectral types for 53 members and 
deep photometric data down to 21 mag in $V$. From time series observations, 
\citet{B07} published a list of variables in the direction of Westerlund 1, but only 
a few cluster members were included in the list. X-ray 
observations opened a new view of Westerlund 1. A magnetar 
(CXO J164710.2-455216), which is a probable member of Westerlund 1, was 
discovered by \citet{M06}. The existence of the magnetar implies that there have been supernova (SN) 
explosions more than once. Many bright X-ray sources have been found from X-ray 
observations \citep{SSZTDP06,CMNDCGG08}. They also discussed the stellar
X-ray properties and X-ray emission mechanisms. Due to the heavy extinction and 
owing to the recent advances in NIR instrumentation, observations in the NIR wavelength 
are favored by many recent studies. \citet{CHCNV06} indentified 24 (including a suspected) 
Wolf-Rayet (WR) stars using NIR photometric and spectroscopic data. According to their 
results, the number of WR stars in Westerlund 1 reaches up to 8\% of all 
known WR stars in the Galaxy. Moreover, a considerable 
number ($\geq 62 \%$) of the stars comprise binary systems. 
The presence of intermediate-mass pre-main sequence (PMS) stars 
was firstly identified with NIR photometry by \citet{BCSWNG08}. 
Unfortunately, \citet{CMNDCGG08} could not find the NIR counterparts of 
several X-ray sources, which were suspected to be low-mass PMS stars in Westerlund 1.
 \citet{MT07} obtained the dynamical mass of $6.3\times10^4 M_{\sun}$ 
using the integrated NIR spectra.

Interestingly, Westerlund 1 contains a substantial number of massive evolved stars at 
various transitional stages. The spectral types of 
these stars were determined in several previous studies \citep{CN02,CN04,CNCG05,CHCNV06,MT07,NCR10,CRN10,DCNJC10}. 
Thanks to the richness of massive transitional objects, it is possible 
to study the evolution of massive stars as well as to observationally 
test their mass loss rate before undergoing an SN explosion. Most 
stars detected in the radio continuum observations do not show a significant mass loss 
that could explain the current mass of Wolf-Rayet stars, while a 
bright radio source, W9, shows a higher mass loss rate and an extended nebula that is indicative of heavy 
mass loss in the past \citep{DCNJC10}. It is suggested that short-lived and 
episodic mass loss events due to unknown mechanisms are required to 
explain the current mass of these evolved stars. Several previous studies have attempted 
to monitor several evolved stars spectroscopically to understand 
the properties of variability as well as to constrain the initial mass of 
the magnetar \citep{RCNC09,RCNN09,RCNL10,CRN10,CRNCDJL11}. Recently, \citet{KB12} determined the fundamental 
parameters of four eclipsing binaries in Westerlund 1.

Despite many detailed studies, the distance to Westerlund 1 is still 
being debated. Early studies, e.g. \citet{W61}, 
\citet{BKS70}, and \citet{PBC98}, favored a closer distance (1.1--1.4 kpc) 
from the brightness of supergiants. \citet{BCSWNG08} 
determined a distance of $3.6\pm0.2$ kpc from NIR photometry and theoretical 
isochrones. A refined analysis by \citet{GBSH11} suggested a distance of 
$4.0\pm0.2$ kpc. \citet{KD07} independently derived a 
distance of $3.9\pm0.7$ kpc based on their new Galactic rotation curve from 
HI observation. \citet{CNCG05}, \citet{CHCNV06}, and \citet{NCR10}  
prefer a distance of 5 kpc which is well supported by their observational data. 
Although the wide range in the distance is now converging to 4.0--5.0 kpc, the exact distance 
to Westerlund 1 remains  the most crucial parameter in the study of Westerlund 1.

The IMF is the most important tool with which to understand star formation 
processes, the formation and evolution of stellar systems, the contribution of 
stellar mass to the host galaxy, the overall stellar properties of 
unresolved remote galaxies, and the evolution of galaxies. In particular, 
the IMF of starburst clusters could give clues to understanding 
the active star forming conditions in starburst galaxies, and therefore 
it is possible to extend this view to the formation stage of the Milky Way. 
Unfortunately, such clusters in the Galaxy are mostly distant and highly 
reddened. Detailed studies are very difficult because of these observational limitations. 
In spite of such difficulties, several studies on the IMF of the starburst 
clusters or well-known active star forming regions have been made. 
The slope of the IMF ($\Gamma$) is -0.7 -- -1.1 for \objectname{the Arches} cluster
\citep{FKMSRM99,KFKN07,ESM09}, -1.3 -- -1.4 for \objectname{R136} 
cluster \citep{MH98}, -0.7 -- -0.9 for \objectname{NGC 3603} 
\citep{EQZG98,SB04,SBBZ06}, -1.3 for \objectname{Trumpler 
14} and \objectname{16} in the $\eta$ Carina nebula \citep{HSB12}, -1.3 
for \objectname{NGC 6530} \citep{SCB00}, and -1.2 for \objectname{NGC 6231} 
\citep{SBL98}, respectively. The slope of the IMF of Westerlund 1 
is known to be -1.3 -- -1.4 \citep{BCSWNG08,GBSH11}. Although 
they statistically derived the IMF of Westerlund 1 from deep NIR data, their data are 
unsuitable for the study of the IMF of massive stars due to saturation of the bright stars. 
Moreover, the masses inferred from NIR photometry is rather uncertain, especially 
for the massive stars, due to the lower resolution of NIR colors.
Fortunately, the starburst cluster 
Westerlund 1 is still accessible for observations in the optical passband with a 4m class telescope, 
and therefore it is possible to study the shape of the IMF of the massive part in detail. 

Open clusters, being stellar systems, are subject to dynamical evolution. 
To understand dynamical evolution in open clusters, many models and 
numerical simulations have focused on mass segregation. As the mass segregation 
is predicted as a consequence of energy equipartition, 
one could expect that open clusters 
younger than the relaxation time are unlikely to show mass segregation. 
However, mass segregation has been found even in young open clusters, 
such as the Orion Nebula Cluster \citep{HH98} and \objectname{NGC 6231} 
\citep{RM98}. Several recent attempts have been made to understand such mass 
segregation in the context of dynamical evolution \citep{MVZ07,MB09,A09,A10}. 
They showed that substructure and subvirial conditions at early stages of 
cluster formation can lead to mass segregation as a consequence of dynamical 
evolution within a very short timescale. On the other hand, 
primordial mass segregation has also been suggested by 
competitive accretion \citep{BD98,BBZ98,BB06} or core coalescence 
\citep{DSGKH08}. In this context, the dynamical evolution of young 
starburst clusters is a very interesting topic because of their 
large numbers of massive stars. A feature indicative of mass 
segregation below $32M_{\sun}$ has been found by \citet{BCSWNG08} who 
showed a difference in the half-mass radii of high- 
and low-mass stars. \citet{GBSH11} supplied strong evidence of mass 
segregation from the spatial variation of the IMF. 
However, any mass segregation among the most massive stars, which can 
cause a violent relaxation, was not studied, because their data was 
confined to stars with masses below $40M_{\sun}$. 

The purpose of this study of Westerlund 1 is to determine a reliable distance, 
reddening, and age of the cluster, as well as to derive its IMF. In addition, 
we will investigate the mass segregation of the most massive stars, 
which dynamically play a great role in the cluster. To achieve our aims, 
optical and NIR photometric data were obtained using 
4m class telescopes. In Section 2, we describe our observations and 
data sets. In Section 3, we present several photometric diagrams and 
determine the fundamental parameters of Westerlund 1, such as the radius, 
reddening, distance, and age. The IMF and mass segregation are 
addressed in Section 4 and 5, respectively. We present discussions on 
several outstanding questions in Section 6. Finally, the summary is in 
Section 7.

\section{OBSERVATIONS}

\subsection{Optical Data}
The observations of Westerlund 1 were carried out on 2009 March 28 
at the Cerro Tololo Inter-American Observatory (CTIO) in Chile. All the images 
were taken with the Mosaic II CCD ($36\arcmin \times 36\arcmin$) attached to the Blanco 4m telescope. The seeing 
was about $0\farcs8$. Harris $BVI_C$ and SDSS $u$ filters were used. The 
angular size of the cluster is so compact that we could place the cluster 
on the best chip (chip 6) among eight chips, where the field of view of an individual chip is about 
$9\arcmin \times 18\arcmin$. Given its severe extinction, special care 
was paid to the exposure times 10s and $150s\times3$ in $I$, $300s\times3$ 
in $V$, $600s\times3$ in $B$, and 200s, 600s, and 1200s in $u$, respectively. 
In order to transform instrumental magnitudes to standard magnitude reliably, many 
standard stars in the Landolt standard fields \citep{Lan92} were observed at a wide range of 
air masses. We made use of IRAF \footnote{Image Reduction and Analysis Facility is 
developed and distributed by the National Optical Astronomy Observatories, 
which is operated by the Association of Universities for Research in 
Astronomy under operative agreement with the National Science Foundation.}
 packages, MSCRED and DAOPHOT, 
to eliminate the instrumental signatures and to carry out PSF photometry. 

\begin{figure}
\epsscale{.80}
\plotone{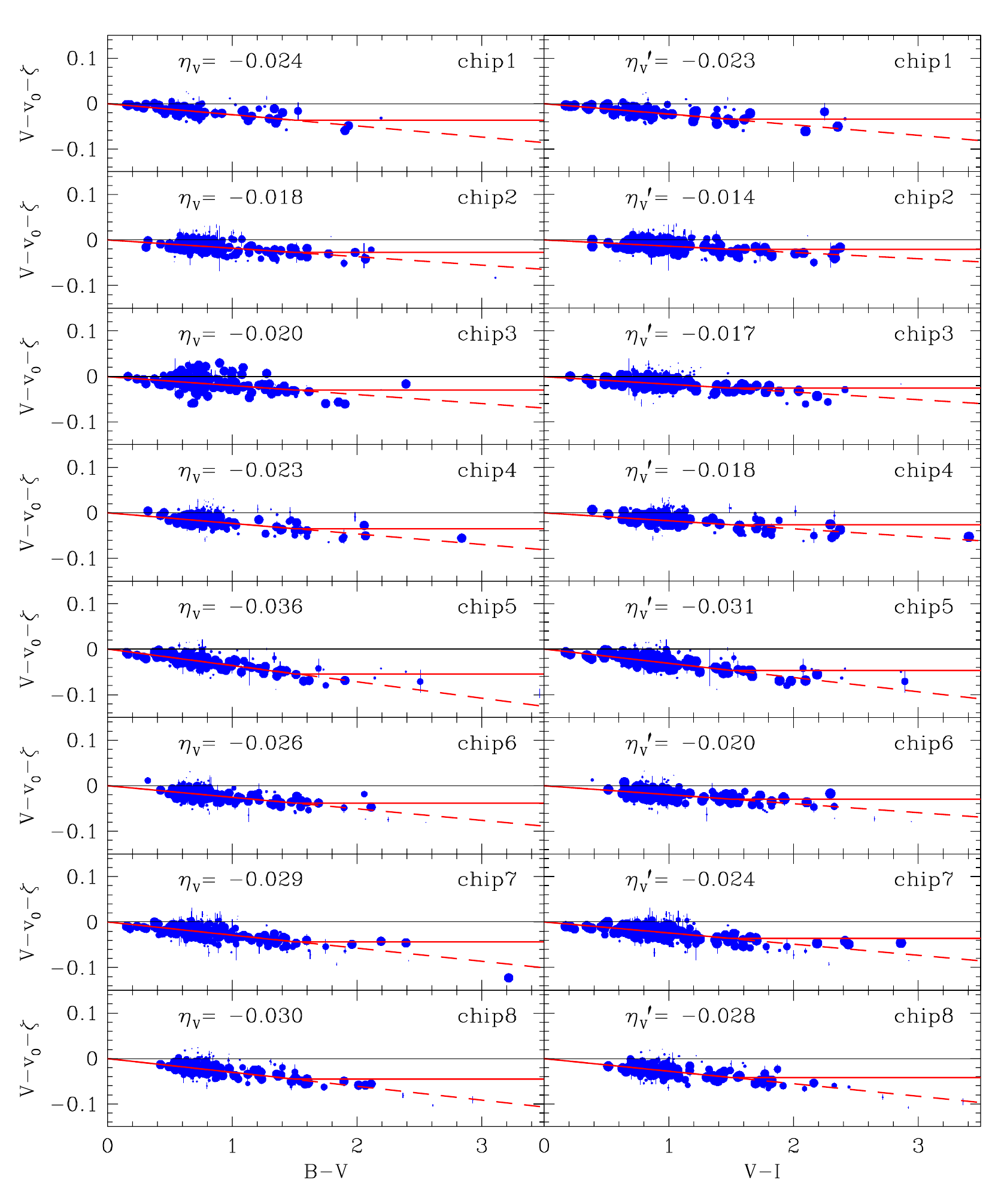}
\caption{The transformation relations of the Mosaic II CCD in the $V$ band with 
respect to $B-V$ and $V-I$. The size of the dots depends on the reliability of a 
given standard star. All the slopes are negative with a break at 1.5 mag in both 
color indices. The details of the transformation relations are addressed 
in the text. We adopt the single slope (dashed line) extrapolated to red colors because 
our main targets are highly reddened OB stars.}
\label{fig1}
\end{figure}

We performed aperture photometry for the standard stars with an aperture 
size of $10 \arcsec$. The instrumental magnitudes were transformed to 
the standard system using the following equation:

\begin{equation}
M_{\lambda} = m_{\lambda}+\eta _{\lambda} \cdot C_0 -(k_{1,\lambda}-k_{2,\lambda} \cdot C_0) \cdot X + \alpha _{\lambda} \cdot \hat{UT} \\
+ \beta _{\lambda} \cdot \hat{x}_{CCD} + \gamma _{\lambda} \cdot \hat{y}_{CCD} + \zeta _{\lambda}
\end{equation}

\noindent where $M_{\lambda}$, $m_{\lambda}$, $\eta _{\lambda}$, $C_0$, $k_{1,\lambda}$, 
$k_{2,\lambda}$, $\alpha _{\lambda}$, and $\zeta _{\lambda}$ represents the 
standard magnitude, the instrumental magnitude, the transformation coefficient, the standard 
color index, the primary extinction coefficient, the secondary extinction coefficient, 
the coefficient for the time variation of photometric zero point, and the photometric 
zero point for a given filter, respectively. The $\beta _{\lambda}$ and $\gamma _{\lambda}$ 
denote the coefficients of spatial variation in magnitude per 1000 pixels 
which will be addressed below. We adopted the Stetson compilations of the Landolt standard star fields 
\citep{st} as the standard system. Using the weighted least square method, 
the atmospheric extinction coefficient, the transformation coefficient, and the 
coefficients for spatial variation were determined. We present the atmospheric extinction 
coefficients of the $BVI_C$ filters for two nights in Table~\ref{tab1}. The transformation 
relations of each band and chip were determined in the form of one or a combination of two straight 
lines as shown in Fig.~\ref{fig1}. In the $V$ transformation, the spectral features due to 
TiO band in late-type stars make the transformation relation, with respective to $V-I$, 
flatter, as shown in Fig.~\ref{fig1} (see also \citep{SB00,LSBKI09}). In the case of Westerlund 1, the single relation 
(dashed line in the figure) extrapolated to extreme red color is adopted because 
the slope in the spectral energy distribution for OB 
stars gradually declines against the reddening effect with wavelength. On the other hand, 
we faced a severe problem with the so-called ``red leak" in the $U$ transformation. Highly 
reddened stars, which may be probable members of Westerlund 1, are too bright and show double-peak 
profiles in the $u$ images due to the red leak of the SDSS $u$ filter. It is 
almost impossible to correct for this amount of red leak due to a lack of standard 
stars with extreme red color. More details are addressed in Appendix A. 

\begin{figure}
\epsscale{.80}
\plotone{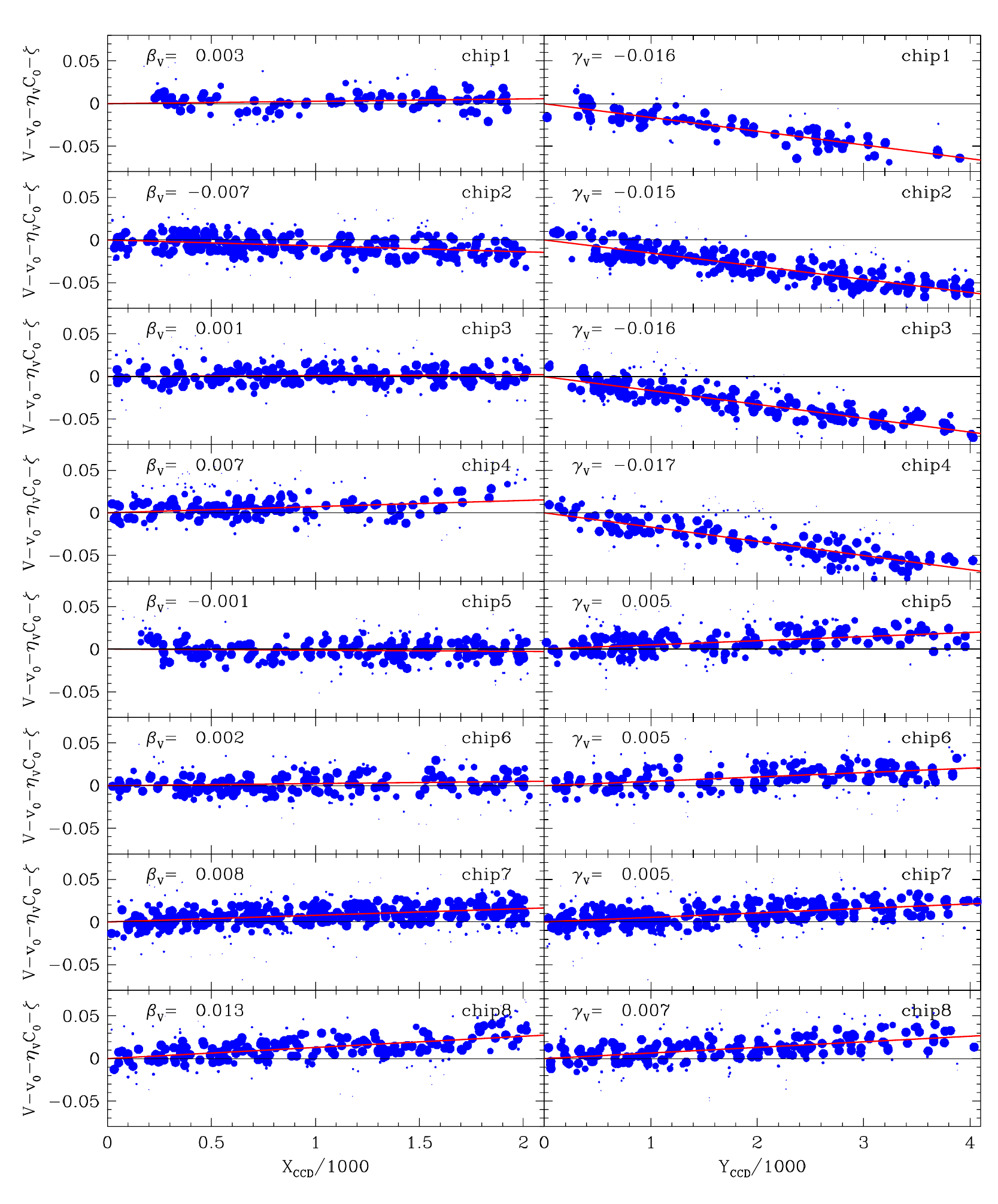}
\caption{The spatial variation of the $V$ magnitude with respect to the $X_{CCD}$ 
(left) or $Y_{CCD}$ (right) coordinates of the Mosaic II CCD. The coordinates of
the standard stars on each CCD chip have been divided 
by 1000 pixel. The variation along $X_{CCD}$ is less significant 
than that along $Y_{CCD}$. Among eight chips, chip 8 shows the largest variation of up to 
0.03 mag along the X axis. The variations for the Y axis are much more prominent in 
the range between -0.07 mag to 0.02 mag. The sign of the slopes is random for the X axis, 
while it is systematic for the Y axis, i.e. negative for chip 1--4 and 
positive for chip 5--8}
\label{fig2}
\end{figure}

Another remarkable feature is a systematic variation in magnitude 
along each axis of a given CCD chip. A similar feature for the CFH12K 
has been reported in \citet{SBCKI08} (see also \citet{MC04}). Fig.~\ref{fig2} represents 
the spatial variations with respect to the $x_{CCD}$ or $y_{CCD}$ 
($\hat{x}_{CCD} \tbond x_{CCD}/1000,~ \hat{y}_{CCD} \tbond y_{CCD}/1000$). The 
pattern of the spatial variation along the $x_{CCD}$ axis seems to be random, 
and the maximum variation is about 0.03 mag for chip 8. On the other hand, the 
variation along the $y_{CCD}$ axis is very conspicuous and systematic. 
The slope $\gamma$ for chips 1--4 is very large (up to 0.07 mag), and is 
larger than that for chips 5--8. This may be related to the spatial variation 
of the focal plane of the Blanco 4m telescope. After correcting for the spatial 
variations, reliable transformation coefficients for each chip could be 
determined. We summarize the transformation coefficients, the coefficients 
for spatial variation, and photometric zero points for all chips in Table.~\ref{tab2}

Recently three optical photometric studies, \citet{PBC98}, \citet{CNCG05}, 
and \citet{B07}, were carried out in the Landolt standard system ($UBVR_C I_C$). 
We compared our data with these previous studies and present the results in Table~\ref{tab3}. 
Although the stars used in the comparison are mostly evolved stars and therefore may be
suspected variables, the photometric data of \citet{PBC98} and \citet{CNCG05} show such
large differences in both $V$ magnitude and color indices, that we attribute the 
discrepancies to their standard transformation. This is particularly so for \citet{PBC98} 
because errors in the extinction coefficients are very  large relative to the values themselves. 
We suspect that they observed Westerlund 1 in  non-photometric conditions or they observed 
only a small number of standard stars. 
\citet{CNCG05}, on the other hand, gave no details concerning their
standard transformations but since their observation was carried 
out in queue mode, it seems likely that an insufficient number of red standard stars were used
to determine the extinction and transformation coefficients. 
The $V$ magnitudes of \citet{B07} are in good agreement with ours, however, 
it is difficult to make a reliable conclusion on the consistency in the colors based on only two or three stars in common.

\begin{figure}
\epsscale{.80}
\plotone{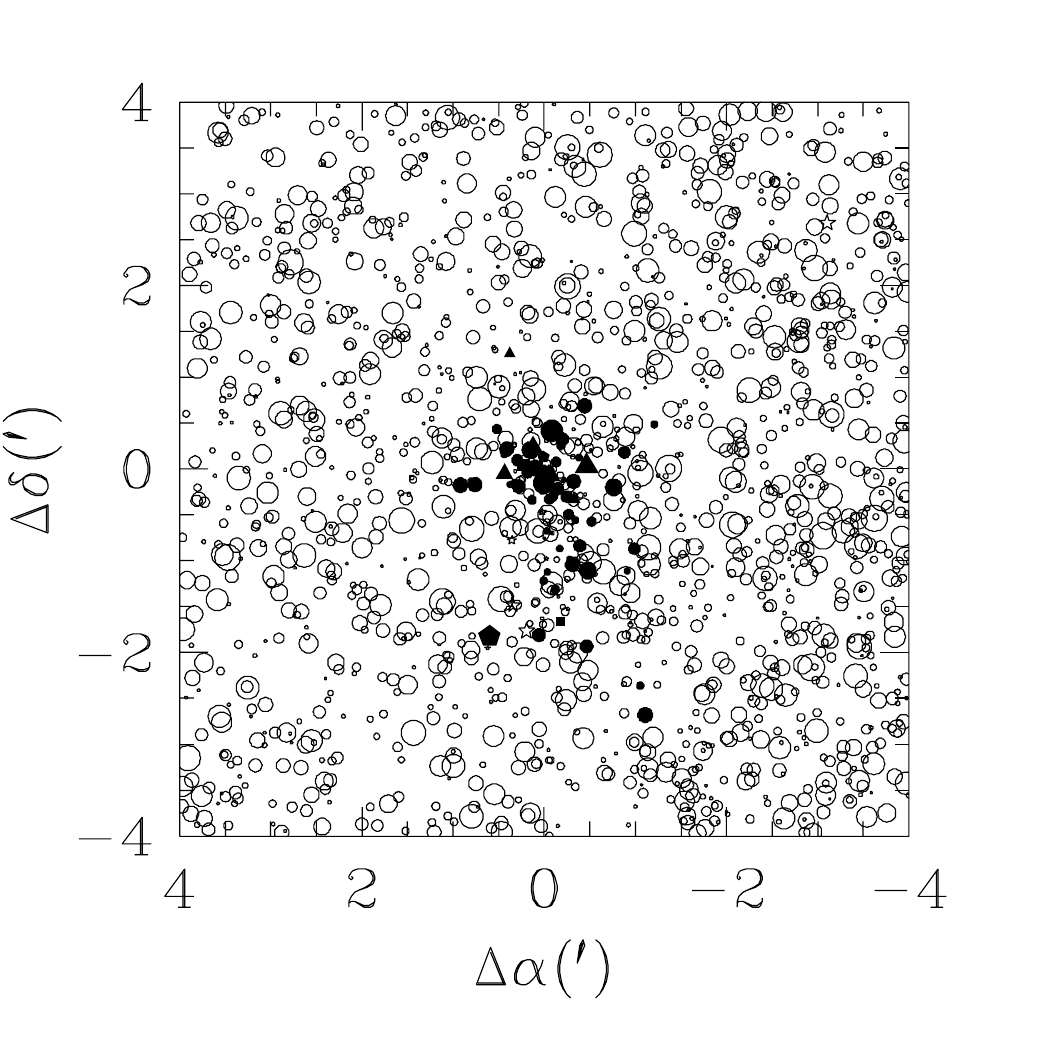}
\caption{The finder chart for Westerlund 1. The filled symbols and star symbols represent the evolved 
stars (dot - OB supergiant, triangle - yellow hypergiant, square - red supergiant, 
and pentagon - luminous blue variable, star symbol - Wolf-Rayet star). The 
size of the symbols is proportional to the brightness, and only stars brighter than 
$V=20.5$ mag are shown. The position of the stars is given relative to the apparent center of the cluster 
($\alpha = 16^h 47^m 4^s.1$ $\delta = -45^{\circ}50\farcm65$, J2000)}
\label{fig3}
\end{figure}

The photometric data for eight bright stars saturated in our data are taken from 
\citet{PBC98} (W32 and W243) and \citet{CNCG05} (W4, W7, W8a, W16a, W33, 
and W57a) after correcting for the differences in photometric zero points. 
We present the finder chart for the stars brighter than $V=20.5$ mag 
in Fig.~\ref{fig3} and photometric data in Table~\ref{tab4}. 
The filled symbols and star symbols in the figure represent 
stars with spectral type and luminosity class in 
\citet{NCR10,CRN10,MT07,CHCNV06}. The evolved stars in transitional stages 
are substantially concentrated in a small area.

\subsection{Near-Infrared Data}

The NIR data were obtained using the IRIS2 detector on the 
Australian Astronomical Telescope (AAT) on 2011 June 25. IRIS2 used a 
Rockwell Science Hawaii-1 HgCdTe 1K infrared detector, covering 
7 \farcm 7 $\times$ 7 \farcm 7 field of view with a $0\farcs45$ 
pixel scale. Two sets of exposure times were used -- long: $10s \times 10$ 
readout $\times 9$ dithering in $J$, $7s \times 10$ readout $\times 9$ 
dithering in $K_S$, and short: $1s \times 10$ readout $\times 9$ 
dithering both $J$ and $K_S$. The basic data reduction was primarily done using 
ORAC-DR in the Starlink package. We subtracted dark and bias from each 
individual dithered image and then normalized them using a flat image. 
The flat image was generated by median combining the dithered images. 
Then the images were cross-correlated and combined to produce the
final image. 

We carried out PSF photometry using DAOPHOT. 
The instrumental magnitudes and colors were transformed to the 2MASS 
photometric system \citep{2mass} using 78 stars in $J$ and 187 
stars in $K_{S}$. As the photometry of bright stars are affected 
by the non-linearity of the IRIS2 detector, and the photometric 
errors of faint stars in the 2MASS catalog are large, the differences 
relative to the 2MASS photometry were calculated only for the stars 
with 11 mag $\le K_{S} \le$ 13 mag. The differences were 
$<\Delta K_{S}> = 0.030 \pm 0.097$ and $<\Delta (J-K_{S})> = 
0.000 \pm 0.122$. Given the errors of the NIR photometry, 
our photometric data are well consistent with the 2MASS system.

\section{PHOTOMETRIC DIAGRAMS}
\begin{figure}
\epsscale{.80}
\plotone{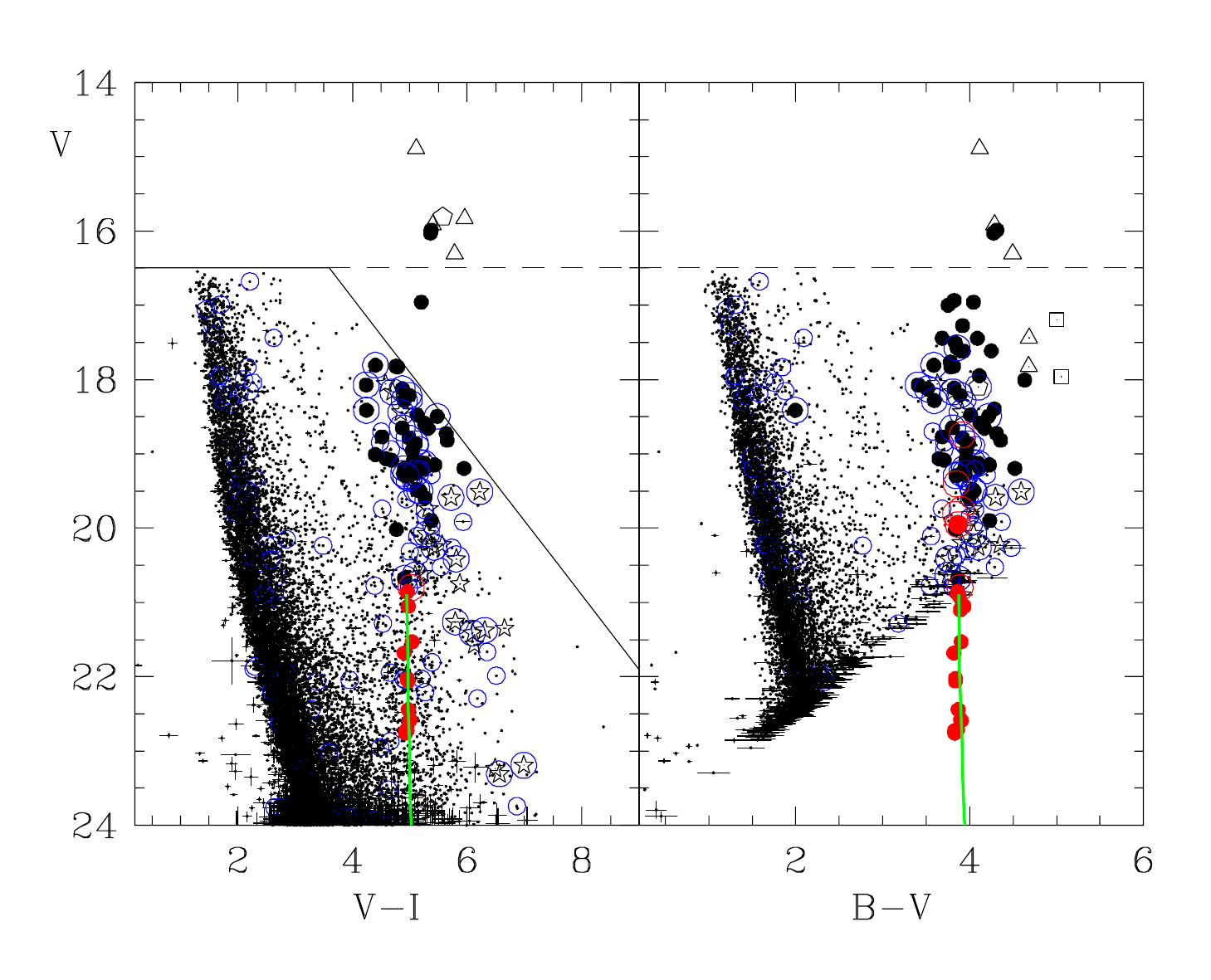}
\caption{The color-magnitude diagrams of Westerlund 1. We present the saturation level in the
$V$ band as a dashed-line, while the upper limit of the $V$ magnitude caused by the saturation 
level in the $I$ band is shown as a solid black line. A large dot, triangle, 
square, open pentagon, and star represent an OB supergiant, a yellow hypergiant, 
a red supergiant, a luminous blue variable, and a Wolf-Rayet star, respectively. A blue 
circle is in addition drawn around the known X-ray emission stars. We overplot the early-type stars (red 
filled circle) and blue supergiants (red open circle) of NGC 6231 as well as the zero-age 
main sequence relation (green solid line) by applying the reddening and distance of 
Westerlund 1, i.e. $<E(B-V)> = 4.19$ mag and $V_{0}-M_{V}=12.9$ mag.}
\label{fig4}
\end{figure}

The highly reddened cluster Westerlund 1 reveals unusual color-magnitude 
diagrams (CMDs) in Fig.~\ref{fig4} due to the presence of many evolved stars, 
such as OB supergiants (OBSGs), yellow hypergiants (YHGs), red
supergiants (RSGs), luminous blue variables (LBVs), and Wolf-Rayet stars (WRs).
In the CMDs, large dots, open triangles, open squares, open 
pentagons, and stars represent OBSGs, YHGs, RSGs, LBVs, and WRs, 
respectively. It would be very useful to compare the stellar content in these 
CMDs with that of other clusters with similar ages. We therefore superimpose
the reddened zero-age main sequence (ZAMS) relation \citep{SB99} 
and the members of \objectname{NGC 6231} \citep{SBL98} 
by adopting the mean reddening and distance modulus of Westerlund 1 determined 
in Section 3.3. The positions of OBSGs of \objectname{NGC 6231} 
are well matched to those of Westerlund 1. The probable main 
sequence (MS) members of Westerlund 1 are barely seen at $V \ge 21$ mag in 
the $(V,V-I)$ diagram. 

\begin{figure}
\epsscale{.90}
\plotone{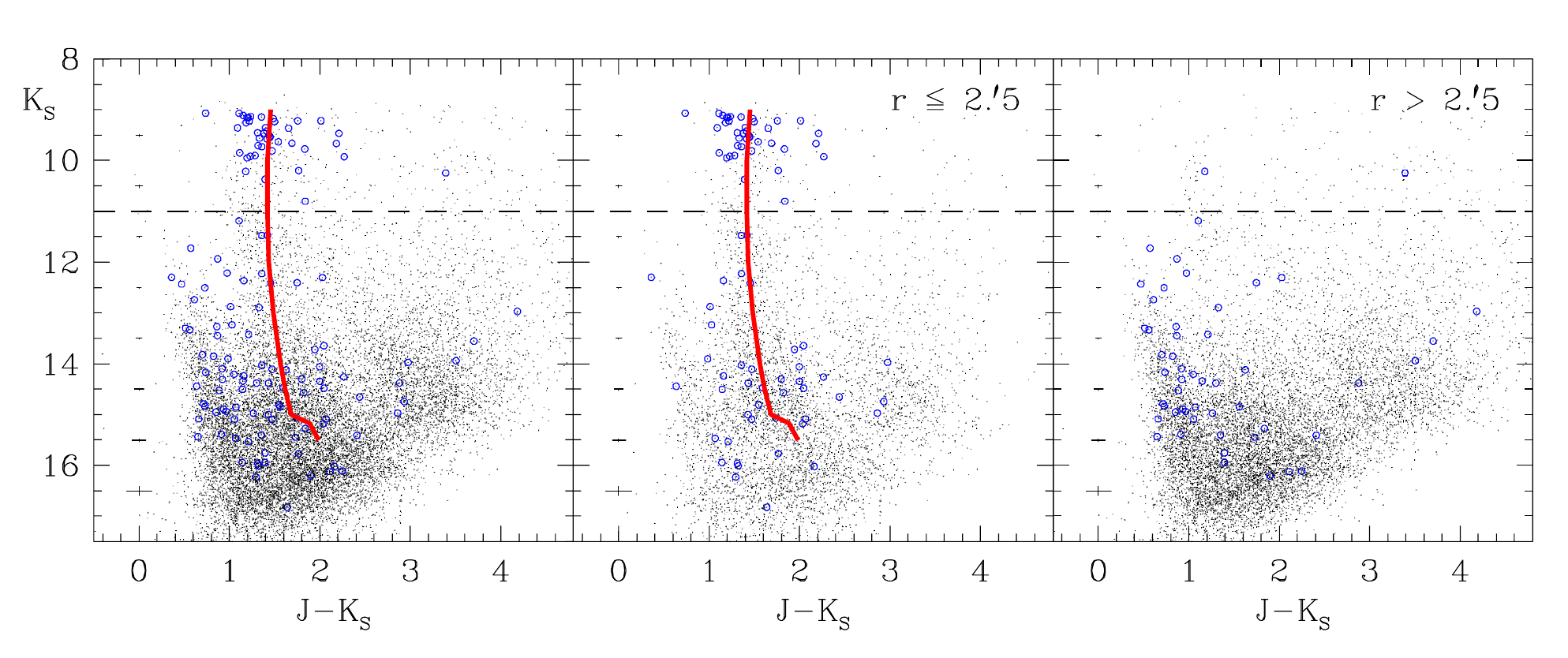}
\caption{The ($K_{S},J-K_{S}$) color-magnitude diagrams for Westerlund 1. 
The left, middle, and right panels show, respectively, all the stars, only the stars within 
$2\farcm5$ from the center to highlight the cluster sequence, and stars 
outside $2\farcm5$. A dashed-line represents 
the linearity limit of the IRIS2 detector. The mean photometric errors are shown 
at the left of all the diagrams. The empirical fiducial line adopted from 
the photometric data of several 
young open clusters is superimposed on the diagrams. The line is shifted 
for a reddening of $<E(J-K_{S})>=1.7$ mag and apparent distance 
modulus of 14.0 mag.}
\label{fig5}
\end{figure}

We present the NIR $(K_{S}, J-K_{S})$ 
CMDs in Fig.~\ref{fig5}. To help distinguish the different sequences of Westerlund 1 stars, we plot in 
the middle panel of the figure, only the stars within a radius of $r = 2.\arcmin5$ from the center.
The foreground stars in the 
Sagittarius-Carina arm constitute the bluest sequence. As shown in 
Figure $4$ of \citet{GBSH11}, the turn-on point from the PMS to the MS is 
found near $K_{S} \sim 15$ mag, and the sequence of MS and evolved 
stars is stretched out within a narrow color range ($J-K_{S} = 1-2$ mag). 
An open circle is drawn around the X-ray sources detected by \citet{CMNDCGG08} 
in Fig.~\ref{fig4} and Fig.~\ref{fig5}. Most of the evolved stars, 
in particular OBSGs and WRs, are X-ray emission objects. 
Since PMS stars are also known as X-ray sources, we attempted to 
trace the PMS locus in Fig.~\ref{fig5} by identifying the 
NIR counterpart of X-ray emission objects. However, only a few X-ray 
emission stars have NIR counterparts in the expected 
PMS locus. We agree with \citet{CMNDCGG08} that many of 
X-ray emitting PMS stars are fainter than $K_S \sim 16$ mag.

In this section, we also present the results from completeness tests and 
determine the fundamental parameters, such as radius, reddening, 
distance, and age.

\subsection{Radius of Westerlund 1}

The surface density profile of a cluster is one of the 
tools used to determine the radius of a cluster as well as to diagnose 
the dynamical structure of the cluster. The profiles also allow us 
to simulate realistically a model cluster regardless of dynamical 
models. In this section, we determine the radius of Westerlund 1 
using the surface density profiles.

\begin{figure}
\epsscale{.40}
\plotone{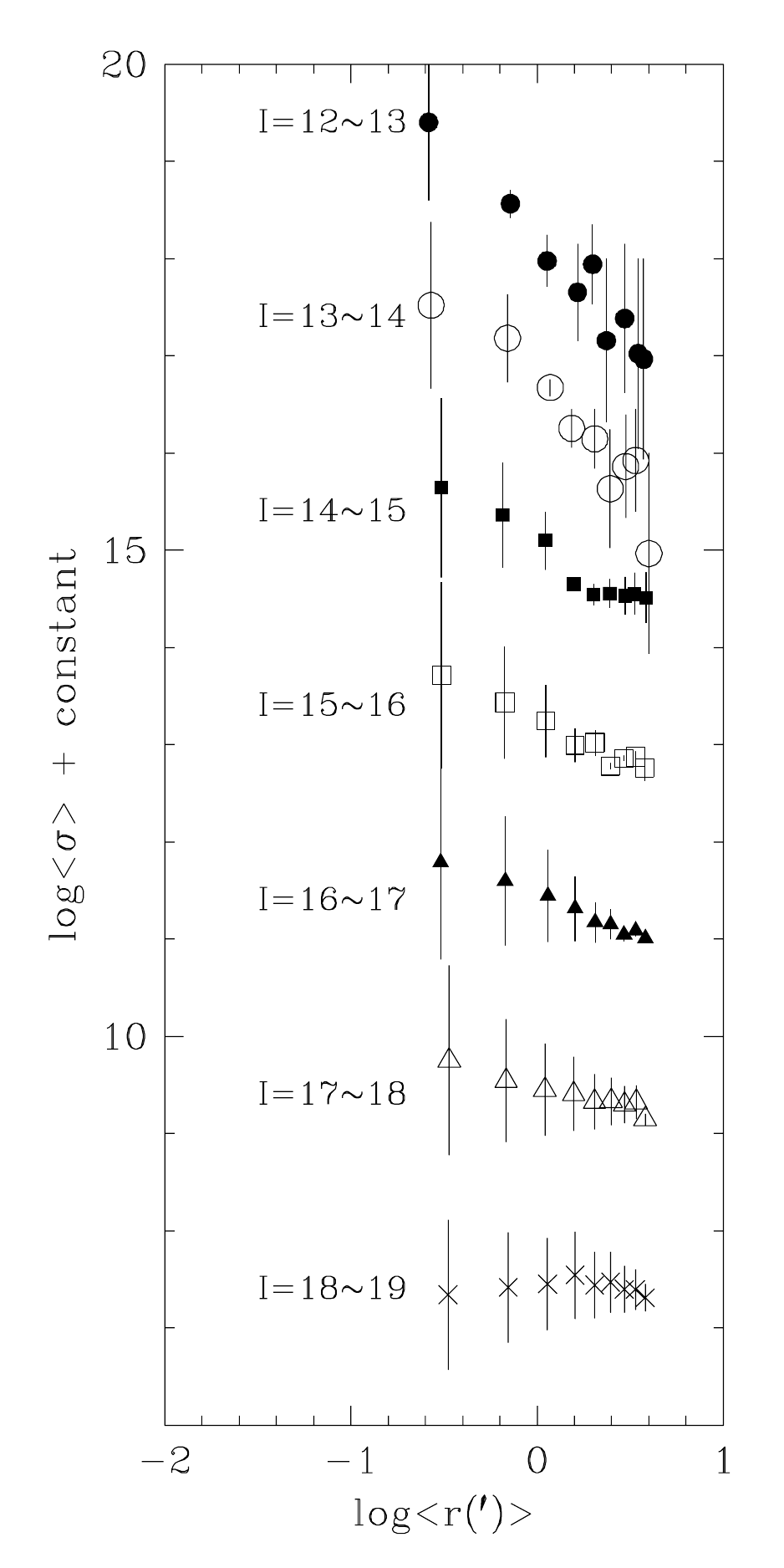}
\caption{The surface density profiles of Westerlund 1. 
The error at a given surface density is assumed to follow Poisson statistics. 
An arbitrary constant has been added to each density profile to separate the plots. 
The surface density of bright stars continues to decrease until 
the level reaches the surface density of field stars at a given brightness. 
The distance from the center to the point where both levels merge is the 
radius of Westerlund 1. We obtain a radius of $2\farcm5$, 
equivalent to $2.8$ pc, for a cluster distance of 3.8 kpc. }
\label{fig6}
\end{figure}

To simplify, we assume a circular shape of Westerlund 1 and assign the densest 
position to the center of the cluster. The width of 
a ring is set to 100 pixels (equivalent to $0\farcm45$), and the 
inner radius of each annuli are increased by 100 pixels. To check 
the difference in surface density profiles with brightness, 
the surface density profiles in the $I$ band are calculated for a 
bin size of 1 mag as in \citet{SBLKL99}. The error in the 
surface density is assumed to follow Poisson statistics, 
i.e. $\epsilon = \pm \sqrt{N}/S$ (where N and S are the number 
of stars and the area of the annulus, respectively). We present the 
surface density profiles in Fig.~\ref{fig6} where an appropriate 
constant is added to each profile to show the difference in profiles 
clearly.  
 
The bright stars are obviously concentrated in the center of the 
cluster. The surface density of bright stars is definitely higher 
in the inner region, while less significant in the outer region. 
The contrast of surface density between the inner and outer regions 
 deceases as the brightness of stars decreases, and finally 
the profiles of faint stars are dominated by Poisson noise. 
We attribute the lack of faint stars in the inner region either to 
the incompleteness of photometry or to dynamical mass segregation, or both. 
The distance from the center to the point where the surface density 
converges to a constant level is about $2\farcm5$ regardless of 
brightness. In order to check the biases from the environmental effect in our 
analysis we have conducted the same procedure for X-ray sources published 
in \citet{CMNDCGG08} as well as our NIR data. Only the X-ray emission stars 
with $V-I > 4.0$ mag were used to avoid non-uniformity of field population 
with X-ray emission. The statistical analysis in different wavelength regions 
gives the same radius as in the optical. It implies that our result may not 
be seriously affected by any environmental effects and that the radius 
derived in our analysis is reliable. We adopt 2.$'$5 as the radius of 
Westerlund 1. This value may be a lower limit.
The physical size is $2.8$ pc if the distance to 
Westerlund 1 is 3.8 kpc (see Section 3.3). The region outside this 
radius is assumed to be the field region.

In addition, we also calculated the half-mass radius of Westerlund 1 using
the radial variation of the IMF (see Section 4) in the mass range of 5 -- 85 
$M_{\sun}$. The half-mass radius of Westerlund 1 is 1.1 pc, which is 39\% of
the adopted radius. This value is consistent with the half-mass radii obtained
by \citet{BCSWNG08} (0.8 pc for m = 10 -- 32 $M_{\sun}$ and 1.1 pc for m = 3.5 
-- 10 $M_{\sun}$). According to their work, the half-mass radius of the low mass 
group is larger than that of the high mass group, and they attributed this to 
the presence of mass segregation. If we take into account the mass segregation
and the unknown distribution of intermediate- to low-mass stars, the half-mass
radius could be larger than the presented value. Hence we also assign the value 
as a lower limit for the half-mass radius of Westerlund 1.

\subsection{Completeness Test}

Completeness of photometry is essential for statistical analyses, 
such as density profiles, luminosity functions, mass functions, 
etc. Therefore, a careful statistical treatment is required because photometric studies may 
be biased toward bright stars. Incompleteness is caused by 
crowding, bright nebulosity, spikes and bright wings of saturated stars, 
etc. For Westerlund 1, the main causes of incompleteness are 
crowding, the spikes of saturated stars, and bright wings, at longer 
wavelengths. \citet{GBSH11} presented the variation in 
the $K_{S}$ magnitude of 50\% completeness across Westerlund 1 using a detailed 2D 
completeness map. In this paper, we check the completeness of our 
optical data.

In order to estimate the completeness of our optical data, we constructed a model 
cluster using a Monte Carlo method. The luminosity function of cluster stars, as well as 
foreground field stars, were derived in the $I$ band because the $I$ magnitude 
is less affected by reddening. Subsequently, the $V$ magnitude and $V-I$ color, with 
appropriate photometric errors, were generated from the $I$ magnitude by taking into account the appearance of the observational CMD. This allowed us to reproduce a realistic cluster. The 
observed radial surface density profiles as shown in Fig.~\ref{fig6} were used to reproduce the crowding 
of the real cluster. In addition, we assumed the radial surface density 
profiles of faint stars to have the same profile as bright ones. On 
the other hand, field stars were uniformly generated across the whole area. 
A set of $V$ and $I$ images were made by adding these artificial stars to each 
sky image. Finally, we carried out PSF 
photometry, aperture correction, and standardization for the artificial 
stars in both the $V$ and $I$ images in the same way as done for the original data. 

\begin{figure}
\epsscale{.80}
\plotone{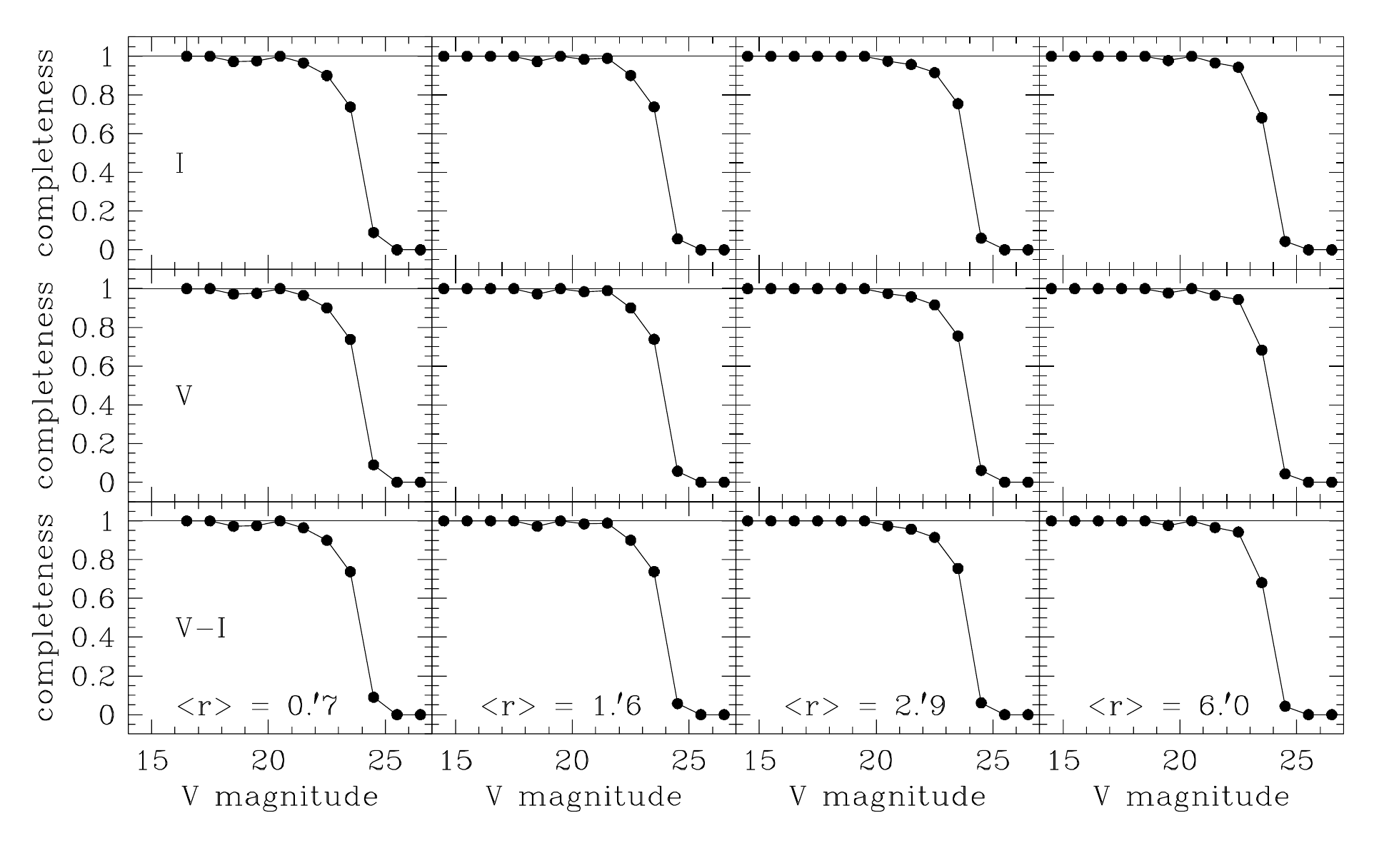}
\caption{The completeness of the photometry against the $V$ magnitude for 
several radii. The completeness in the inner region 
is slightly lower than that of outer region. Our optical photometry 
seems to be about 90 \% complete down to $V\sim 22.0$ mag, even at the cluster center.}
\label{fig7}
\end{figure}

\begin{figure}
\epsscale{.80}
\plotone{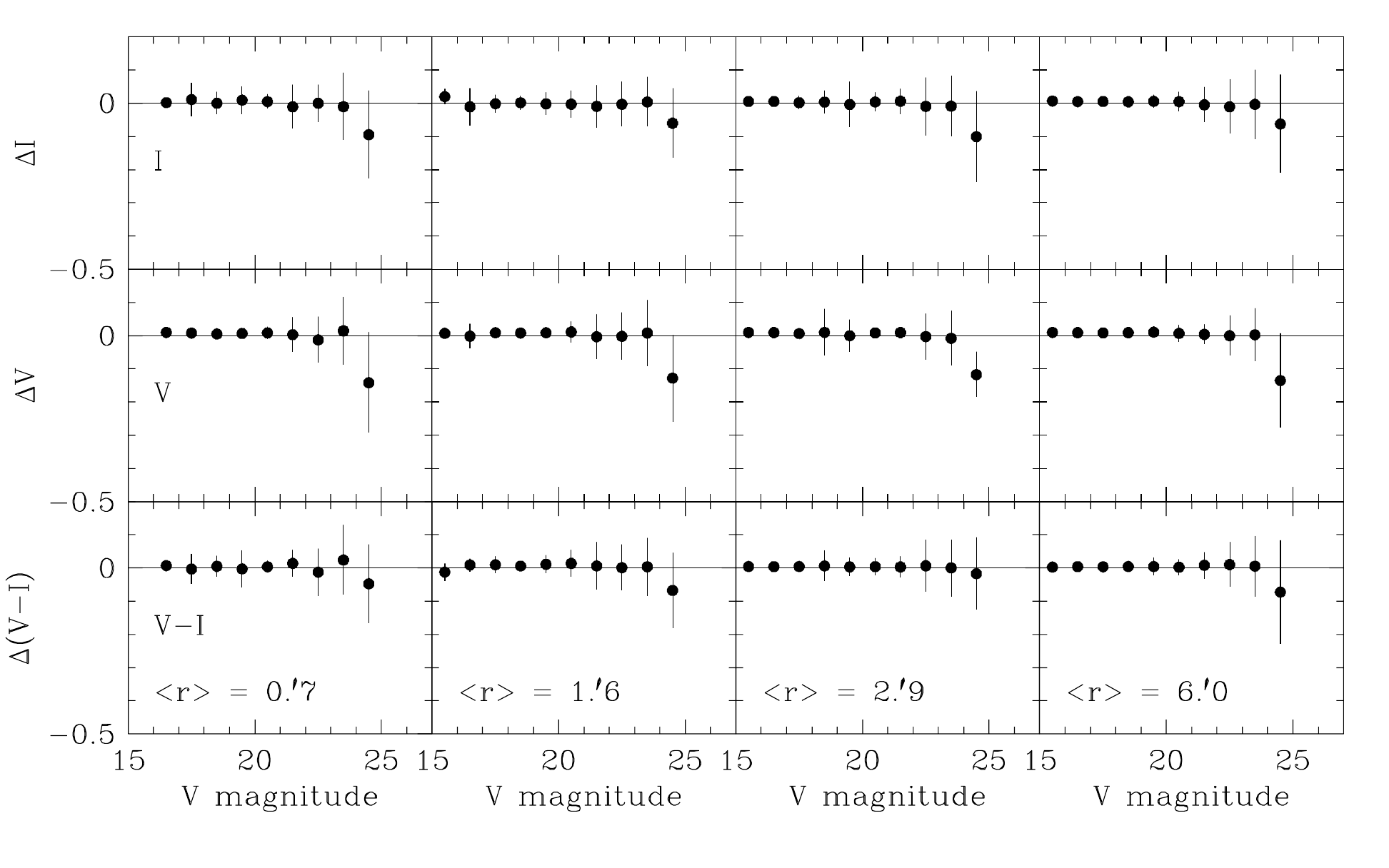}
\caption{The difference between the input and output 
data for several radii. The output magnitudes are 
systematically brighter for faint stars ($V \geq 23.0$ mag.), while 
well consistent with the input magnitudes for the others.}
\label{fig8}
\end{figure}

From the analysis of the model cluster, we estimated the completeness of 
our photometry. Fig.~\ref{fig7} shows the completeness of photometry at different $V$ 
magnitude limits for several annuli. The radial variation of completeness is not as 
strong a function as that found by \citet{GBSH11}. This is related to the observed passband. 
The effect of reddening is much more severe in the optical than in the NIR, and the difference 
in brightness between massive stars and low-mass stars is much larger in the optical, consequently, 
the 90 \% completeness level occurs at $V \sim 22$ mag, equivalent 
to $\sim 20 M_{\sun}$, and may be an upper limit because 
faint stars could be expected to be much affected by the spikes and bright wings of saturated stars 
in long-exposure $I$ band images since we could not reproduce these spikes 
reliably. Moreover, quite a few stars in the $I$ band have no counterpart in 
the $B$ or $V$ bands because of the lower signal in $B$ or $V$ due to the severe 
reddening. Fig.~\ref{fig8} represents the mean difference and the 
standard deviation between the input and output data, 
where $\Delta \equiv$ output - input. The data are well consistent 
with each other for bright stars, while the output data becomes
systematically brighter for $V \geq 23$ mag. This feature from PSF 
photometry using IRAF/DAOPHOT had already been reported in 
\citet{SBLKL99}. The fluctuations in the differences in the 
central region are slightly larger than those in the outer region 
due to the effect of crowding. From this analysis we conclude that our 
optical data are quantitively and qualitatively complete down to 
22 mag in $V$, even at the cluster center.

\subsection{Reddening and Distance} 

In general, the reddening $E(B-V)$ of early type stars is best determined 
from the $(U-B,B-V)$ diagram. Unfortunately, as we could not obtain a 
reliable $U-B$ color index even for the brightest star due to the red leak 
of the SDSS $u$ filter (see Appendix A for details), as well as to the very high 
reddening, a more fundamental, but less reliable method, namely, the spectral type versus intrinsic 
color relation, was used to determine the reddening. Although the 
observation was made only for longer wavelengths ($\lambda > 5500 \AA$) 
to achieve a reasonable signal-to-noise ratio, the spectral classification 
of \citet{NCR10} for OBSGs is the most recent result with a 8.0 m class 
telescope, and was helpful for this work. To determine the color 
excess of individual stars, we adopted the calibration of \citet{f77}, 
which provides the intrinsic $B-V$ color, effective temperature, 
and bolometric correction for various luminosity classes. The intrinsic 
$B-V$ color of 53 OBSGs was determined by interpolating the spectral 
type versus intrinsic relation, and then calculating the reddening $E(B-V)$. 
We obtained the mean value of $<E(B-V)>=4.19\pm0.23$ (s.d.) mag. Assuming 
$R_{V} = 3.1$, the total mean extinction ($A_V$) is about $13.0$ mag. If we simply assume 
the intrinsic value of $B-V = -0.25$ mag and $V-I = -0.28$ mag 
for OBSGs, the ratio of color excess is $<E(V-I)> \over <E(B-V)>$ = 1.26. 
This implies that the reddening law toward Westerlund 1 is fairly normal, i.e. $R_V \sim 3.1$. 

\begin{figure}
\epsscale{.80}
\plotone{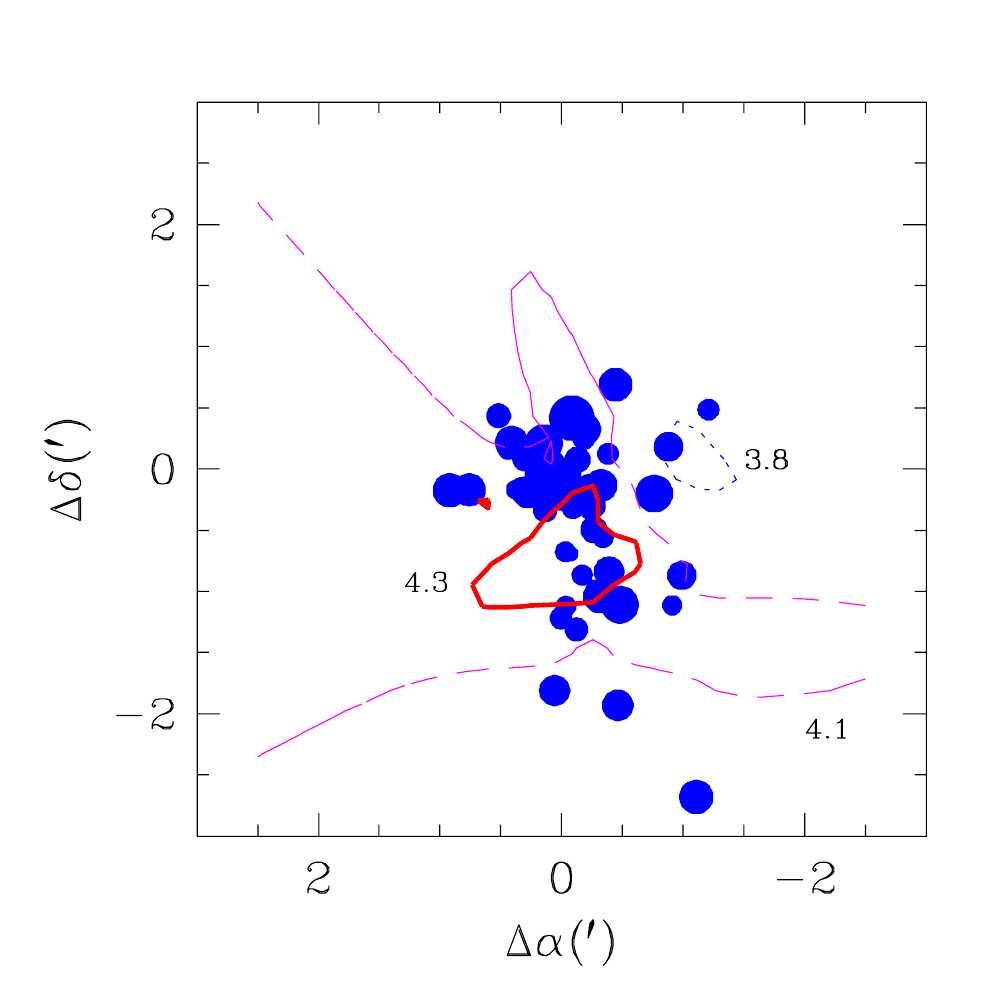}
\caption{The reddening map of Westerlund 1. The blue dots  
represent OB type supergiants, the size of the dots are proportional to 
the brightness. The red solid line, dashed magenta line, and 
dotted blue line, represent contours of $E(B-V)=4.3$, 4.1, and 3.8, 
respectively. The reddening map indicates that the innermost 
region is the most heavily reddened.}
\label{fig9}
\end{figure}

We also constructed the reddening map and present it in Fig.~\ref{fig9}. 
In the figure, the large dots denote the position of 53 OBSGs. And 
solid red, dashed magenta, and dotted blue lines represent contours of $E(B-V)$
= 4.3, 4.1, and 3.8, respectively. The size of the dots are proportional 
to the brightness of individual stars. The reddening is the highest near 
the center and shows a gradient from east to west. In general, we can expect 
a cavity of extinction near cluster center as is seen in NGC 3603 \citep{SB04}
due to the strong stellar wind from massive stars. Contrary to our expectations,
the reddening was highest near the cluster center. The region is coincident
with the extended radio emission source A2 of \citet{DCNJC10} and the 8 $\mu m$
{\it Spitzer} IRAC emission feature (see Fig 4 of \citet{DCNJC10}). According
to \citet{DCNJC10}, the extended radio emission source A2 may be associated
with the WN7 star WR B. These facts imply that outflowing material from WR stars
are forming dust around stars, and is evidenced from the reddening map.

Given photometric differences among previous studies, the results are 
in good agreement with each other within the errors, e.g. $E(B-V)= 4.2 \pm 0.4$ 
\citep{NCR10} and $E(B-V)=4.3\pm0.2$ \citep{PBC98}. \citet{NCR10} 
discussed the large deviation of their average color excess ratio $<E(V-I)> 
\over <E(B-V)>$ from the standard value \citep{RL85}. They attributed
the deviation to the error in their $B$ band photometry, putatively the 
zero point in the band. However, since their photometric data were transformed 
to the standard $R_{C}I_{C}$ system of \citet{Lan92}, the color 
excess ratio should have been compared with the standard reddening 
law in the Johnson-Cousins photometric system, not in the Johnson's 
$RI$ system. Given the error of the color excess ratio of \citet{NCR10}, 
their result seems to be close to the standard reddening law.

The distance to open clusters is, in general, 
determined by fitting the ZAMS relation to the MS of the cluster. 
However Westerlund 1 is so obscured that it is nearly impossible to define 
the MS of Westerlund 1 in the ($V$, $V-I$) CMD. Although the location of 
MS stars in the CMD is well-defined, the color range of early-type 
stars is very narrow, and the MS stars appear to be almost 
perpendicular to the color axis in the CMD. And therefore it 
would give a less reliable distance to the cluster if the ZAMS 
relation were employed. \citet{PBC98} seems to be confronted 
with this difficulty. They determined a distance to Westerlund 1 of 
1.5 -- 2.0 kpc by matching the absolute magnitudes of 
supergiants to the isochrone of age 4.0 Myr. However, given the uncertainties 
in the reddening correction of individual stars and the intrinsic 
scatter of evolved stars (see Fig 3. in \citet{HD79}), it seems 
that the distance obtained by \citet{PBC98} is less reliable. 
Indeed the value is substantially different from those of other 
estimates, e.g. 5.0 kpc -- \citet{CNCG05,NCR10,CHCNV06}, 3.9 kpc 
-- \citet{KD07}, and 4.0 kpc -- \citet{GBSH11,KB12}.

For this reason, the use of NIR photometric data, which is less 
affected by interstellar extinction, is inevitable in determining 
the distance to Westerlund 1. As shown in both \citet{BCSWNG08} and 
\citet{GBSH11}, the appearance of the turn-on point from the PMS to the
MS (they called it the transition zone) provides a fiducial marker for the
empirical fitting method and encourages us to construct an empirical 
relation using photometric data of other clusters. \citet{MNLSJ07} have
constructed empirical isochrones in the ($V, V-I$) diagram to determine
fundamental parameters of young stellar associations and clusters. Since the 
turn-on position is sensitive to the age, we had to select clusters with 
almost the same age as Westerlund 1, $\sim 5.0$ Myr.
The young open clusters \objectname{NGC 2362} (3.0--6.0 Myr -- \citet{M01}), 
\objectname{NGC 6231} (2.5--4.0 Myr -- \citet{SBL98}; 3.0--5.0 Myr 
-- \citet{BVF99}; 4.5 Myr -- \citet{SB93}), and \objectname{NGC 6823} 
(2.0--7.0 Myr -- \citet{BBD08}) were selected under this condition. 
The empirical fiducial line (EFL) constructed in the $[M_{K_{S}},
(J-K_{S})_0]$ plane is presented in Fig.~\ref{fig5} after correcting for
the reddening and the difference in distance. The EFL is very useful 
because it is independent of stellar evolution models as well as being
less affected by variations in metallicity. Details on 
the EFL are addressed in Appendix B. We brightened the ($K_{S},J-K_{S}$) 
CMD of Westerlund 1 by an apparent distance modulus of 14.0 mag to fit it to 
the turn-on point from the PMS to the MS of the EFL by eye and compared 
the observed color indices with those of EFL for stars with 
$1 < J-K_{S} < 2$ and $11 < K_{S} < 13$. The mean color excess 
of $<E(J-K_{S})> = 1.70\pm0.21$ was thus obtained. In order to 
compute the total extinction in $K_{S}$, i.e. $A_{K_{S}}$, equation 
2 of \citet{KFL06} was used, where we adopted the exponent 
of $\alpha = 1.66$ according to their suggestion. The reddening 
law results in $A_{K_{S}} = 1.12\pm0.14$ mag. Consequently, the 
apparent distance modulus subtracted by $A_{K_{S}}$ is $K_{S0}-M_{K_{S}} 
= 12.9$ mag, equivalent 
to 3.8 kpc. Our result is consistent with the three previous studies 
\citep{KD07,GBSH11,KB12}. 

The distance to Westerlund 1 is still being debating in previous studies. 
\citet{CNCG05}, \citet{NCR10}, and \citet{CHCNV06} prefer a distance of 5 kpc  
which places Westerlund 1 in the Norma arm. On the other hand, a distance of 1.1 kpc 
adopted by \citet{PBC98} corresponds to the Sagittarius arm. 
To constrain the distance of the cluster we referred to the NIR CMD of control 
field in Figure 4 (upper right panel) of \citet{GBSH11} and assumed that the 
Sagittarius arm, Scutum-Centaurus arm, and Norma arm to be at 
$1.1\pm0.4$ kpc, $3.0\pm1.0$ kpc, and $5.4\pm0.8$ kpc toward Westerlund 1 
($l=339\,^{\circ}.5, b=-0\,^{\circ}.4$), respectively. If Westerlund 1 is 
in the Sagittarius arm, the bluest sequence in Figure 4 of \citet{GBSH11} 
should penetrate the sequence of Westerlund 1 or slightly overlap it because 
they are a co-spatial disk population. As there is an obvious gap below the MS 
turn-on of Westerlund 1 (see also the upper left panel in their figure), 
the distance derivations of \citet{W61,BKS70,PBC98} should be ruled out. 
It can be assumed that the group of stars at $K_{S}\sim14$ mag and 
$J-K_{S} \sim 0.8$ are the representative field star population of the spiral 
arm and have the same photometric properties in all three arms, and 
that the effects of reddening within a single spiral arm are negligible. 
The bright stars, which are likely to belong to the Sagittarius arm, will be 
at $K_{S}\sim17.3$ mag and $J-K_{S}\sim2.5$ if those are reddened and dimmed 
by the reddening of Westerlund 1 [$E(J-K_S) = 1.7$ mag and $A_{Ks} = 1.1$ mag]
and the difference in distance moduli [$\Delta (K_{S0}-M_{K_{S}}) =2.2$ mag] 
between the Sagittarius arm and Scutum-Centaurus arm. This approach 
allows us to explain the presence of a population with $2 \lesssim J-K_{S} 
\lesssim 3$. 

On the other hand, the bright stars would be shifted to $K_{S}\sim18.6$ 
mag were the stars in the Norma arm, i.e. $\Delta(K_{S0}-M_{K_{S}}) =3.5$ 
mag instead of 2.2 mag. Were Westerlund 1 in the Norma arm, the field star 
population in the arm should have barely been detected because it would approach 
the detection limit of $K_{S}$. See Figure 4 (upper right panel) 
of \citet{GBSH11} again, the population of field giants is obliquely 
stretched over a wide color range ($1 \lesssim J-K_{S} \lesssim 5$). 
There seems to be a void near $K_{S} \sim 12$ mag and $J-K_{S}\sim2.2$ mag 
in the sequence. If this feature is real, it implies that there is 
an interarm region beyond the Sagittarius arm in the line of sight, and 
that the redder sequence of field giants belongs to the second spiral arm 
from the Sun toward Westerlund 1. In this way, the stars with $J-K_{S} \ge 2.0$ 
in the CMD are likely to be a population in the Scutum-Centaurus arm. 
In conclusion, we believe we can well constrain the location of Westerlund 1 to the 
Scutum-Centaurus arm which is consistent with our result, and make 
a thread of connection between our interpretation and the results of 
\citet{KD07}.

\subsection{Age of Westerlund 1}
The widely used method for estimating the age of an open cluster is to fit the 
theoretical isochrone to the MS turn-off in the CMDs or the Hertzsprung-Russell 
diagram (HRD), by assuming that the members are coeval. For the case of 
Westerlund 1, since the MS turn-off is very faint due to severe extinction and 
the MS turn-on is too faint to be detected from optical photometry it is nearly 
impossible to directly apply this method to the optical CMDs. For this reason, 
previous studies have constrained the age of Westerlund 1 to be 3--5 Myr 
using the lifetime of massive evolved stars \citep{CNCG05,CHCNV06}, 
isochrone fitting to the evolved stars \citep{NCR10}, or the PMS-MS 
transition zone in the NIR CMD \citep{BCSWNG08,GBSH11}. Recently \citet{KBGR12}
have conducted a Bayesian analysis using kinematic members. In the case of 
\citet{NCR10}, their adopted isochrone seems not to be the best prediction for 
the positions of evolved stars in the HRD, especially O type supergiants. Since 
the difference in the temperature scales between \citet{MSH05} and \citet{f77} 
for late-O type supergiants is not very significant, we attribute their results 
to the incorrect estimate of the distance to Westerlund 1. \citet{sbl97,SCB00} found 
a discrepancy in the age and mass of PMS stars among PMS evolution models. 
\citet{NCR10} also pointed out a younger age of Westerlund 1 inferred from PMS 
model \citep{BCSWNG08}. It implies that the evolution models of PMS stars 
still have some limitations, e.g. the introduction and handling of mass
accretion in PMS evolution models \citep{hosokawa11}. \citet{KBGR12} did not take into account the 
comprehensive member selection for RSGs and X-ray emitting PMS stars, whose 
membership is very important to draw the formation picture of 
Westerlund 1. We would like to suggest that more careful member selection 
may be needed in their work. In this paper, for the evolved stars, 
we will use new theoretical stellar evolution models with/without 
rotation \citep{rot12} and $Z=0.014$ to estimate the age of Westerlund 1 
in the same manner as \citet{NCR10} and compare the result with that of 
previous studies. 

\begin{figure}
\epsscale{.80}
\plotone{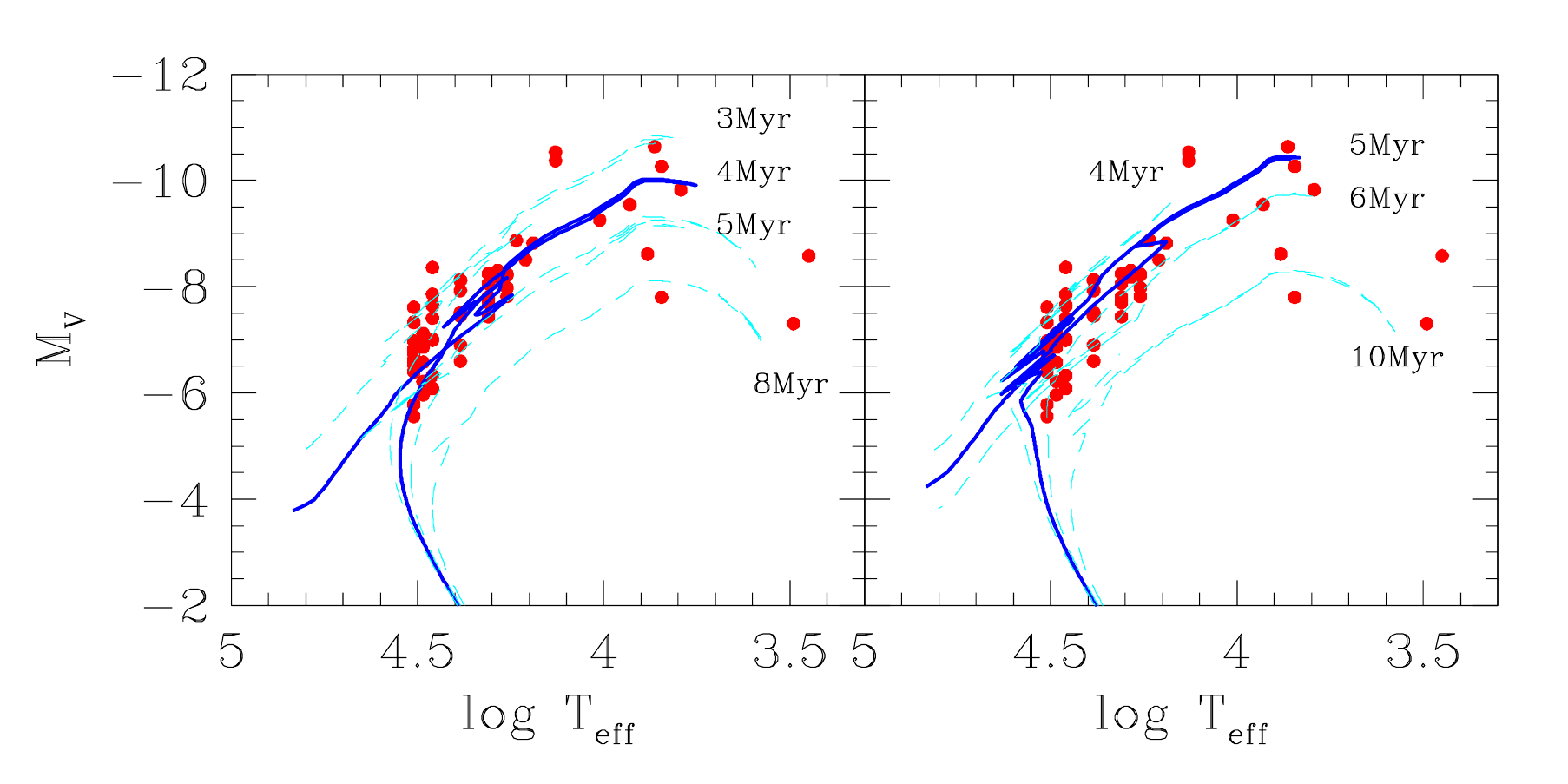}
\caption{The age of Westerlund 1. The left panel shows the loci of stellar 
evolution models without rotation and the right panel the loci of models with rotation. 
Each red dot represents 62 evolved stars, and four 
isochrones with different age are superimposed on each 
diagram. The solid blue line denotes the best fit model. 
The stellar evolution models without rotation give a 
slightly younger age than those with
rotation. The latter models seem to reproduce very well 
the positions of the evolved stars in the Hertzsprung-Russell diagram. }
\label{fig10}
\end{figure}

We applied the total extinction inferred from OBSGs and the distance modulus 
of 12.9 mag to the 62 evolved stars in Westerlund 1. The spectral type of 
individual stars was transformed to the effective temperature using 
the calibration of \citet{f77}. The observational HRD of the evolved 
stars is shown in Fig.~ \ref{fig10}. Finally, four isochrones of 
different ages are superimposed on the diagrams. The 4.0 Myr 
isochrone (blue solid line) without rotation gives the best fit to the evolved 
stars in the left panel of Fig.~\ref{fig10}. The model well predicts the 
position of YHGs in the HRD, however, the predicted position of OBSGs 
is slightly cooler than observed. Most of the observed stars are fainter 
and cooler than predicted, when the 3.0 Myr isochrone 
(upper dashed line) is placed on the HRD. The age of 3.5 Myr seems 
to be matched well to our data. Indeed, this age is the same as 
that estimated from the stellar evolution models of \citet{SSMM92}. 
On the other hand, the stellar models with rotation seem 
to better predict the position of observed stars in the HRD (see right 
panel of Fig.~\ref{fig10}). The 5.0 Myr isochrone (blue solid line) 
traces not only the locus of the  OBSGs but also those of 
the two late-B supergiants and five bright YHGs. Most of the massive 
stars are confined between the 4.0 Myr and 6.0 Myr isochrones, 
respectively. Although the high binary fraction of $\ge 62\%$ 
\citep{CHCNV06}, variability \citep{B07,CRN10}, and uncertainties in 
the spectral classification \citep{NCR10} may affect the location 
of the evolved stars in the HRD, we assign the age of Westerlund 1 to be 
5.0 $\pm$ 1.0 Myr, if we assume a near coeval star formation (see 
also \citet{KBGR12}).

Our result of $5.0 \pm 1.0$ Myr from NIR data is compatible with the results of 
\citet{CNCG05,CHCNV06,NCR10}, which discussed the evolutionary timescale of 
massive stars, and with that of \citet{KBGR12} based on a Bayesian analysis as 
well as with that of \citet{GBSH11} based on the isochrone fitting. It is 
also consistent with the formerly assumed age of Westerlund 1 (3 -- 6 Myr) in 
Section 3.3. However, the presence of RSGs and relatively faint YHGs makes us 
concerned as to whether or not the population within Westerlund 1 is coeval. 
If these two groups of stars are real members, current evolution models 
indicate that the age of Westerlund 1 should be extended to 8.0 -- 10.0 Myr 
as shown in Fig.~\ref{fig10}. More discussions will be presented in Section 6.1.

\section{THE INITIAL MASS FUNCTION}

Using the relation between the spectral type and bolometric correction 
of \citet{f77}, the absolute magnitude of the evolved stars were
converted to the absolute bolometric magnitude. Next, we constructed a 
mass-luminosity relation, especially for the stars at the post-MS stage, 
based on the probability distribution of a given mass along the 
evolutionary track. We directly interpolated the stellar evolution 
models and determined the masses of individual stars from their positions 
in the HRD. However, the masses of LBVs and WRs still remain uncertain because 
the intrinsic properties of these stars are still not well-known. 
\citet{CHCNV06} inferred that the initial masses of WRs in Westerlund 1 is 
in the range of 40 -- 55$M_{\sun}$ from the evolutionary models. Assuming 
the Kroupa IMF, a simple simulation was conducted to estimate the mass range 
of WRs using the stellar evolution models \citep{rot12}. The masses of model 
WN and WC stars are distributed in the range of 50--70$M_{\sun}$ and 40 --
70$M_{\sun}$, respectively. Since it is impossible to determine the exact 
initial mass of individual WRs from the current data and current calibrations, 
all WR stars were included in the highest mass bin. On the other hand, 
according to \citet{M03}, only the most masssive stars with an initial mass 
exceeding $85M_{\sun}$ could evolve to become LBVs. The masses of the two LBVs 
(W9 and W243) in Westerlund 1 may be at least $85M_{\sun}$. The two LBVs were 
also included in the highest mass bin. For the rest of the stars, which are 
presumably MS or stars near the MS turn-off, i.e. those with $V-I \geq 4$ and 
$B-V \geq 3$ and which lie within the radius 
of $2\farcm5$, the $V$ magnitude was converted to absolute magnitude using 
the reddening map in Fig.~\ref{fig9} and a distance modulus 
of 12.9 mag. For these stars, the conversion from color indices 
to the effective temperature is much more uncertain unless their
spectral type or individual reddening is available, and 
therefore the mass-absolute magnitude relation of the 5.0 Myr isochrone 
with rotation was used in order to determine their masses. 
We also determined the mass of field interlopers, with the same 
photometric properties as Westerlund 1, to subtract the contribution of 
field stars in the mass function, after correcting for the 
mean extinction and distance modulus. Note that two YHGs 
(W12a and W265), four RSGs (W20, W26, W75, and W237), 
two OBSGs (W228b and W373), and four WRs (N, T, X, and S) 
have been excluded in the mass function calculation because the membership 
of these stars is uncertain, and because these OB and WR stars are outside 
the adopted radius of Westerlund 1 ($r_{cl} = 2\farcm5$).
Recent simulations on the dynamical evolution of young open clusters \citep{A10}
suggested the existence of ejected stars due to the interaction between massive 
stars. In addition, as Westerlund 1 may have experienced a number of
supernovae, we could expect a number of runaway stars around Westerlund 1.
We therefore cannot exclude the possibility that these massive stars may
be ejected members of Westerlund 1. The membership of these stars is left
for future study.

To calculate the mass function $\xi$ ($\equiv N/\Delta\log m/$ area),
we counted, respectively, the number of stars within the adopted radius
of Westerlund 1 and field region ($r > 2.'5$) for a given mass bin 
($\Delta \log m = 0.2$). The latter was normalized by multiplying the areal ratio 
of cluster to field. This therefore represents the probable number density of 
field interlopers within the cluster region. The total number density of stars 
within each mass bin in the cluster region was subsequently adjusted
by subtracting the number of interlopers in the same mass bin. A 
complementary IMF for the stars with intermediate-mass was 
also derived from NIR photometry. As the model provides $M_V$ and $(B-V)_0$, 
the absolute $V$ magnitude of  the 5 Myr isochrone with rotation, was converted into 
the absolute $K_S$ magnitude using the ($V-K_S,B-V$) relation calculated 
by one of the authors (M.S.B.). The transformed $M_{K_{S}}$ was dimmed by 
the apparent distance modulus of 14 mag and transformed to mass for a given 
star, by interpolating the observed $K_S$ magnitude into the mass-luminosity
relation, where only MS stars which were not affected by the non-linearity 
of the NIR detector were used. The stars away from the cluster region 
were also used to estimate the contribution to the mass function by field 
interlopers. A rather reliable IMF in the low-mass regime was obtained by 
\citet{GBSH11}. We present the IMF of Westerlund 1 in Fig.~\ref{fig11}. 
To avoid effects of binning, an additional mass function was calculated by 
shifting the mass bin by $0.1$ in logarithmic mass scale. The 
large dots and open circles represents the IMFs derived from optical data 
and NIR data, respectively. The error is assumed to follow Poisson statistics.
According to the completeness of 
our optical photometry, the reliable mass limit from the optical data 
is about $20M_{\sun}$, and the IMF below this limit may be a lower limit. 
The shape of the IMF is very similar to that of \objectname{NGC 3603} (the 
dashed line in the figure). The IMF revealed a shallow slope ($\Gamma = -0.8
\pm 0.1$), when we computed the slope in the mass range of log $m = 0.7$ to 
log $m = 2$. If we exclude the two highest mass bins because of the loss 
of massive stars through evolution, the slope of the IMF is much shallower 
($\Gamma = -0.5$). Although there are some uncertainties in the IMF derived 
from NIR photometry, such as the differential reddening, incompleteness, 
and subtraction of field interlopers, the slope is in good agreement with 
the results of \citet{SB04} and \citet{SBBZ06} ($\Gamma = -0.9$) for 
\objectname{NGC 3603}. The shape and slope of the IMF, $\Gamma=-0.7\pm0.1$,  
did not vary significantly when we included the excluded stars mentioned 
above (two YHGs, four RSGs, two OBSGs, and four WRs). More details about 
the shape of the IMF will be discussed in Section 6.2.

Observational evidence of a large number  of low-mass stars present 
in starburst open clusters has been reported. Based on this evidence, 
we computed the total mass of Westerlund 1 assuming the IMF of the intermediate- 
to low-mass regime ($10 M_{\sun} > m > 0.08 M_{\sun}$) to be that of 
\objectname{NGC 2264} \citep{SB10} and that of \citet{K02}, respectively. 
After appropriately scaling the level of the two IMFs to ours and multiplying by
the surface area and the mass bin, the IMFs in the massive to low-mass regime 
were integrated. The total mass of Westerlund 1 is about $7.3 \times 10^4 
M_{\sun}$ from the IMF (\objectname{NGC 2264} plus ours) and about $7.8 
\times 10^4 M_{\sun}$ from that of Kroupa plus ours, respectively. 
These are well consistent with the dynamical mass of \citet{MT07} within 
their uncertainty. The total mass derived by \citet{GBSH11} would be 
comparable to ours, had they adopted a rather flat slope in the massive 
part of the IMF, instead of a single overall slope.

Another way to estimate the total mass of Westerlund 1 is to use the information
on the radial variation of the IMF in Fig.~\ref{fig12} (see section 5 for 
details). Due to a lack of information on the radial distribution of 
intermediate- to low-mass stars, a single slope of the IMF extrapolated 
to 5 $M_{\sun}$ was assumed within a given annulus. The sum of masses derived 
within five annuli yielded a total mass of $5.1 \times 10^4 M_{\sun}$. 
This value is for masses greater than 5 M$_\odot$, and therefore is 
a lower limit to the total mass of Westerlund 1. Again, we confirm
that Westerlund 1 is the most massive open cluster in the Galaxy by comparing 
the total mass of other starburst open clusters (see Table 4 in \citet{CNCG05}).
Hence, the mass of Westerlund 1 is comparable to that of globular clusters 
in the Galaxy, and the study of Westerlund 1 could give hints to 
the formation process of globular clusters in the early Universe.

\section{MASS SEGREGATION}

Observational evidence of mass segregation in open clusters has been 
confirmed from the radial variations of mean mass, the ratio of high-mass 
to low-mass, the slope of the IMF, and minimal spanning tree method. 
In this paper, we investigated the mass segregation using the radial 
variation of the IMF of bright evolved stars. From the masses of individual 
stars determined above, the IMFs were computed for different annuli 
of width of $0\farcm5$. The IMF for a given annulus was adjusted 
for the contribution of field interlopers. The slopes of the 
IMF were computed only for $\log m \geq 1.4$. We present the 
result in Fig.~\ref{fig12} (left panel). 

Obviously, the IMF of the innermost region shows the shallowest 
slope. The slopes gradually steepen with increasing distance from the 
center, however, there is an abrupt change at the fourth ring 
($r=1\farcm7$). If this feature is not related to a statistical fluctuation, 
it may imply a substructure in Westerlund 1. Unlike \citet{GBSH11}, we simply 
assumed a circular shape for the cluster. The departure from circularity 
could cause a distorted surface density profile. Although 
an apparent elongation is evident within $r=1\farcm4$, the slopes of the 
IMF become systematically steeper within the annulus. We also performed 
the same approach as above by taking into account the elongation using the same 
parameters as in \citet{GBSH11}, $a:b = 3:2$ and a tilt angle $\sim 
22.^{\circ}5$ (see Fig. \ref{fig12} right panel). 
The semi-major axis is increased by $0\farcm 5$. The slopes of the IMF were 
systematically steeper ($\Gamma =$ -1.3, -1.6, -1.7, and -3.2) as 
the central distance increases within $a = 2\farcm 0$, while a relatively
flat slope ($\Gamma =$ -1.8) was found between $a = 2\farcm0$ and 
$a = 2\farcm 5$. Although the discontinuity in the slopes occurs 
at a slightly different position with the different approaches, 
our result suggests the presence of a halo 
structure in Westerlund 1. To check whether or not a similar pattern could be 
found in the surface density profiles, we drew surface density profiles for 
several cut-off magnitude in Fig.~\ref{fig13}. A slight change in the slope 
could be seen at $\log r\sim 0.2$, equivalent to $1\farcm6$ in the surface 
density profiles. Since the same trend could be found in all profiles with
different cut-off $I$ magnitudes, the change in slope may not be 
a consequence of data incompleteness. If the abrupt change in the radial 
variation of the IMFs is related to that of the surface density profile, 
it may also imply that Westerlund 1 has substructure (a core and a halo). 
Although we could not confirm the existence of a substructure in Westerlund 1, 
exploring the presence of a halo based on reliable membership criteria, will 
be one of the important challenges to understand the dynamical 
evolution of massive open clusters. \citet{KSB10} found a halo in the 
surface density profile of a compact young cluster \objectname{Hogg 15}.

On the other hand, there may be another interpretation for the fluctuation 
of the slope of the IMF within a certain radius, namely, limited mass 
segregation. Mass segregation was only found in the inner region, while massive 
stars in the outer region were well-mixed. According to \citet{RM98} 
and \citet{HH98}, mass segregation occurs only for massive stars 
in \objectname{NGC 6231} and the \objectname{Orion Nebula 
Cluster}. However, \citet{GBSH11} found mass segregation even for 
the low-mass stars in Westerlund 1, and therefore the result may not be a
mass segregation that is limited to a particular mass range. Although the 
fluctuation is rather large, we view the trend in the radial variation of 
the IMFs as evidence of mass segregation. But since statistical uncertainties 
may dominate in the outer region due to the small number of massive stars, 
it is hard to measure the degree of mass segregation reliably. Violent 
processes among massive stars, such as dynamical ejection, mass loss from 
evolved stars, and SN explosions may also affect the trend in the radial 
variation. 

\section{DISCUSSION}

\subsection{The Presence of the Unexpectedly Faint Evolved Stars}
As mentioned in Section 3.4, the presence of relatively faint YHGs (W12a and 
W265) and RSGs (W20, W26, W75, and W237) is confusing. Owing to good seeing 
conditions, we could resolve W12a into two stars in the optical images. 
The star is about 1 mag fainter than the expected brightness from the stellar 
evolution model. According to \citet{B07}, W12a, W265, W20, and W237 did 
not reveal photometric variability during seventeen nights. Neither could 
\citet{CRN10} find any spectroscopic variability. However, \citet{RCNC09} 
found photospheric pulsation in W265 and \citet{CRN10} found variation 
in the spectral types of the RSGs. In our data, W20 was barely detected 
in the $V$ band, whereas the star was saturated in the $I$ band. For the case 
of W75, the star was only detected in the $I$ band. Given the limiting 
magnitude ($V =24$ mag) of our $V$ band data, W20 and W75 were much fainter 
than other RSGs (W26 and W237, $V \sim 17.5$ mag). We also checked the 
brightness of the stars in the $K$ magnitude of \citet{BKS70}. Obviously, 
W20 and W75 were about $\sim$ 1 mag fainter than W26 or W237 in the $K$ band. 

As all previous studies noted, if the stars are physically connected to 
Westerlund 1, two issues need to be clarified. The first is whether or 
not differences in brightness are possibly due to variability. 
The Galactic YHG \objectname{V509 Cas} reveals photometric variations from 
4.89 mag to 5.18 mag in the $V$ band \citep{PZ92}. The star gradually dimmed 
from 1980 to 1990 with, in addition, short-term fluctuations. An episodic 
variation of a Galactic YHG, \objectname{$\rho$ Cas}, in the $V$ band 
has also been reported by \citet{L03}. The brightness suddenly 
dropped by $1.3\pm0.1$ mag in 2000. Therefore, the brightness 
of faint YHGs could be explained by variability. On the other hand, 
there are significant differences in the brightness among four RSGs as 
mentioned above. Since most RSGs are long period variables 
(LPVs), the difference between faint (W20 and W75) and bright 
(W26 and W237) RSGs could be explained by pulsation if the stars are coeval. 
According to \citet{KSB06}, the brightness of the extreme supergiant LPV 
\objectname{VX Sgr} varies by up to 5.8 mag in the $V$ band. If W20 and W75 
belong to this extreme class, the stars could be as bright as other RSGs 
in Westerlund 1 at maximum light. However, such an extreme class is very rare. 
Furthermore, the classification and the evolutionary status of \objectname{
VX Sgr} is still uncertain as mentioned in \citet{CLMD10}. \citet{WBF83} have 
studied the properties of a sample of LPV supergiants in the Small and Large Magellanic 
Clouds. They found a smaller mean pulsation amplitude of $\leq$ 0.25 mag in $K$ for 
RSGs. Given the small variations of RSGs from the large sample, the difference 
of $\sim$ 1 mag in $K$ among four RSGs in Westerlund 1 may not be explained 
by intrinsic variability, hence, cluster membership of the two faint RSGs 
(W20 and W75) may be ruled out. To confirm this, photometric monitoring 
would be very useful. 

The other issue is the presence of RSGs and the age of the 
cluster ($5.0 \pm 1.0$ Myr). According to the stellar evolution 
model with rotation \citep{rot12}, the RSGs with mass of 
$25 M_{\sun}$ are found at an age of 8 Myr.  If they are cluster members,
the age spread of  Westerlund 1, is much larger than that 
discussed in Section 3.4. In the case of PMS stars in 
young open clusters, an age spread of $\Delta \tau_{age,PMS} \sim 4.0$--
$6.0$ Myr is a typical value, e.g. \citet{SCB00,PSK01}, and 
\citet{PS02}. There are several uncertainties, such as differential 
reddening, the presence of accretion disks, star formation
history within a cluster \citep{SB10}, particularly the dependency on the 
stellar evolution models or PMS evolution models. Unfortunately, 
the age spread of massive stars are poorly known due to several 
factors, such as rarity, high frequency of multiplicity, and the effect of 
rotation on the stellar evolution, etc. Consequently, 
if the age spread could be as large as that of PMS stars, Westerlund 1 
might be formed by a single starburst but with a large age spread. 
We also note that uncertainties in the stellar evolution 
models of massive stars cannot be excluded. On the other hand, 
\citet{KBGR12} have found a smaller age spread ($\Delta \tau \leq 0.4$ Myr) 
for Westerlund 1, while \citet{SB04}, \citet{PPG11}, and \citet{RBSGGR10} found 
clues to a large age spread in another starburst cluster 
\objectname{NGC 3603}. The result of \citet{KBGR12} seems statistically 
inadequate to conclude  that the age spread in Westerlund 1 is very small
because they used photometric data for a very narrow field of view. 
If their argument is correct, the faint evolved stars should be excluded 
from membership, or some other astrophysical interpretations are required
to explain the range of stellar ages. 
 
One possible scenario is related to a rather complicated history 
in the earliest epochs of star formation in Westerlund 1 as evidenced by two 
important observations. \citet{RCNL10} attempted to constrain the initial 
mass of the progenitor of the magnetar (CXO J164710.2-455216) from photometric 
and spectroscopic studies of the eclipsing binary W13 and suggested that 
the progenitor mass was larger than $45 M_{\sun}$. In the second observation, 
\cite{KD07} found a large bubble with a minimum dynamical age of 2.5 Myr in 
Westerlund 1, and suggested that the bubble was a sign of very early massive 
star evolution in the cluster. However, adopting a higher inclination angle 
would give a larger physical size for the bubble and subsequently a larger 
dynamical age of up to 5.0 Myr. Given the large age spread, the presence of 
a magnetar, and the dynamical age of the expanding bubble, significant star 
formation events may have occurred twice in Westerlund 1. The difference in 
the age of the RSGs ($\tau \geq 8.0$ Myr) and Westerlund 1 ($\tau = 5.0 \pm 
1.0$ Myr) corresponds to the lifetime of the most massive star with a mass 
of $120 M_{\sun}$. The progenitors of RSGs and the relatively faint YHGs may 
have been formed together with the most massive stars in the giant molecular 
cloud. After 3.0--4.0 Myr, the most massive star would have exploded as an SNe, 
and the hot gas ejected from the SNe explosion could be the bubble found by 
\citet{KD07}. The explosion may have induced the formation of second 
generation stars in the rest of 
the natal cloud. The current population of Westerlund 1 may have
formed at this epoch. A similar star formation history had been suggested 
by \citet{SB04} in another starburst cluster \objectname{NGC 3603}. 
The authors found an age difference between early 
type stars in the core and halo, which implies two different 
star formation events. Indeed, \citet{RBSGGR10} found the presence of 
PMS stars with a relatively old age from their proper motion study. 
The results by \citet{KBGR12} (see their Fig. 2) and \citet{PPG11} 
also suggested the presence of an older main sequence population 
both in the fields around the clusters Westerlund 1 and \objectname{NGC 3603}. 
To confirm such a star formation history in Westerlund 1 it will be necessary 
to study the low-mass stars in the halo of Westerlund 1 that are still in 
the PMS stage. Perhaps, confirmation of this scenario may give 
important information on the formation of starburst clusters. On the other 
hand, \citet{GBSH11} argued that the apparent elongation and mass 
segregation of the cluster could be explained by merging of 
multiple coeval subclusters in the same natal giant molecular cloud 
in the early phase of formation, given no evidence for a large age spread.  

If the relatively low luminosities of the several evolved stars, 
W12a, W20, W26, W237, and W265, are not related to 
variability, we attribute the dimming to differential reddening. 
According to \citet{LD74}, the total extinction 
of W20 ($A_V =16.5$ mag) and W75 ($A_V = 19.2$ mag) is higher 
than those of W26 ($A_V=12.1$ mag) and W237 ($A_V=13.6$ mag). 
Internal reddening by the extended material itself could 
cause such a difference in brightness among the RSGs.
On the other hand, in the 8.6-GHz images 
of \citet{DCNJC10}, W20, W26, W12a, and W265 reveal 
significant cometary shapes pointing toward the anti-center of the 
cluster, while W75 and W237 show no such asymmetrical structure. 
According to current stellar evolution models, the existence of 
such stars cannot be explained by a single burst of star formation 
5 Myr ago. Indeed, only a few RSGs can be statistically 
predicted from our simple simulations -- a cluster with an age spread 
of 3 Myr and the IMF similar to that of \objectname{NGC 3603} or 
an age spread of 5 Myr and the Kroupa IMF. However, with such limited 
data we could not find strong evidence for the membership of these stars. 
Consequently, establishing or refuting, the membership of such 
stars will provide important clues to understanding the 
formation history of massive young clusters.

\subsection{The Shape of the Initial Mass Function}

\begin{figure}
\epsscale{.80}
\plotone{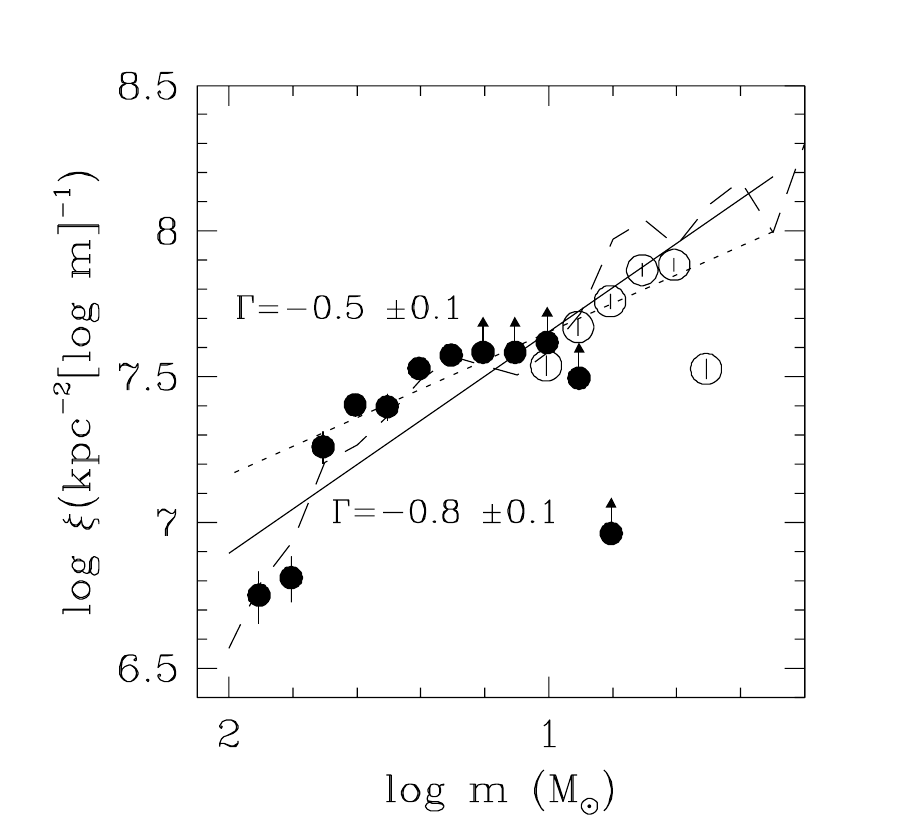}
\caption{The initial mass function of Westerlund 1. The dots
and circles represent the initial mass function derived from 
optical photometry and near-infrared photometry, respectively. The dashed 
line denotes the scaled initial mass function of NGC 3603 \citep{SB04}. 
We obtain a shallow slope of $\Gamma = -0.8\pm 0.1$ in the mass range of 
$\log m =$ 0.7 -- 2.0. The shape of the IMF of NGC 3603 and Westerlund 1 
are very similar.}
\label{fig11}
\end{figure}

\begin{figure}
\epsscale{.80}
\plottwo{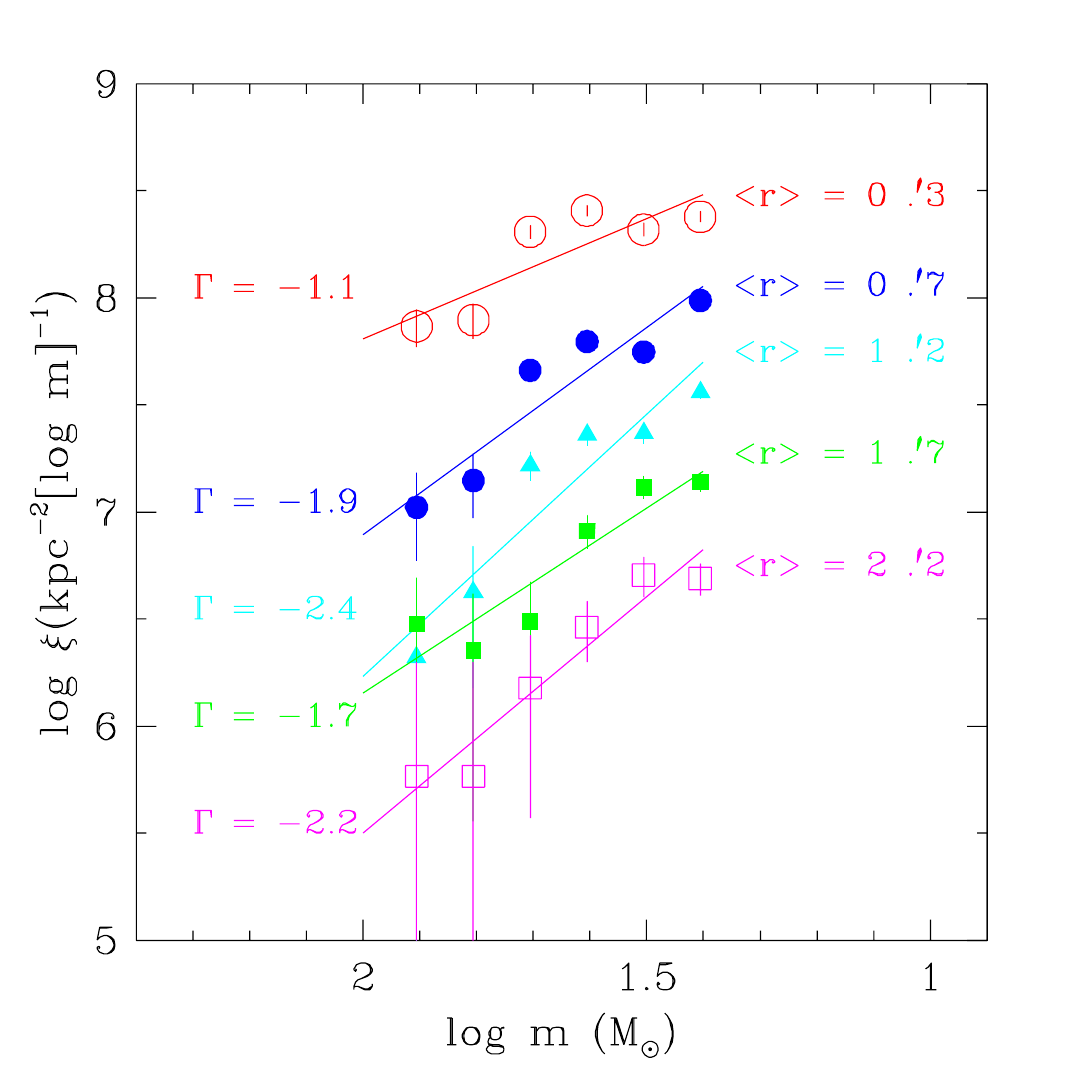}{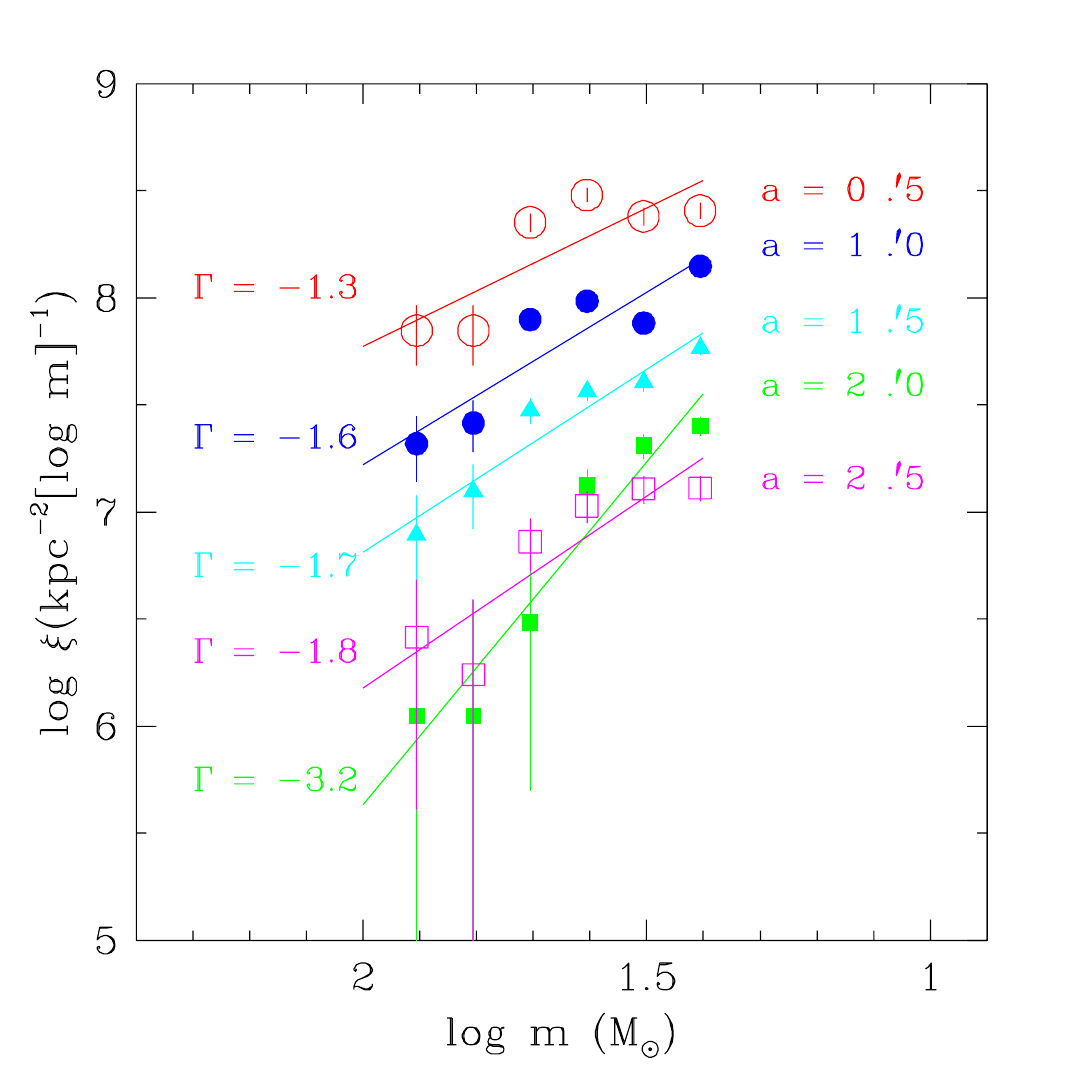}
\caption{The radial variation of the IMF. We present 
the IMFs measured at increasing distance from cluster center from top to bottom 
in the diagrams. (Left) circular fitting, (Right) ellipse fitting.
The slopes become systematically steeper inside $r\sim1\farcm7$ or
$a \sim 2.'0$ where an abrupt change in slope occurs. See the main text 
for details.}
\label{fig12}
\end{figure}

\begin{figure}
\epsscale{.80}
\plotone{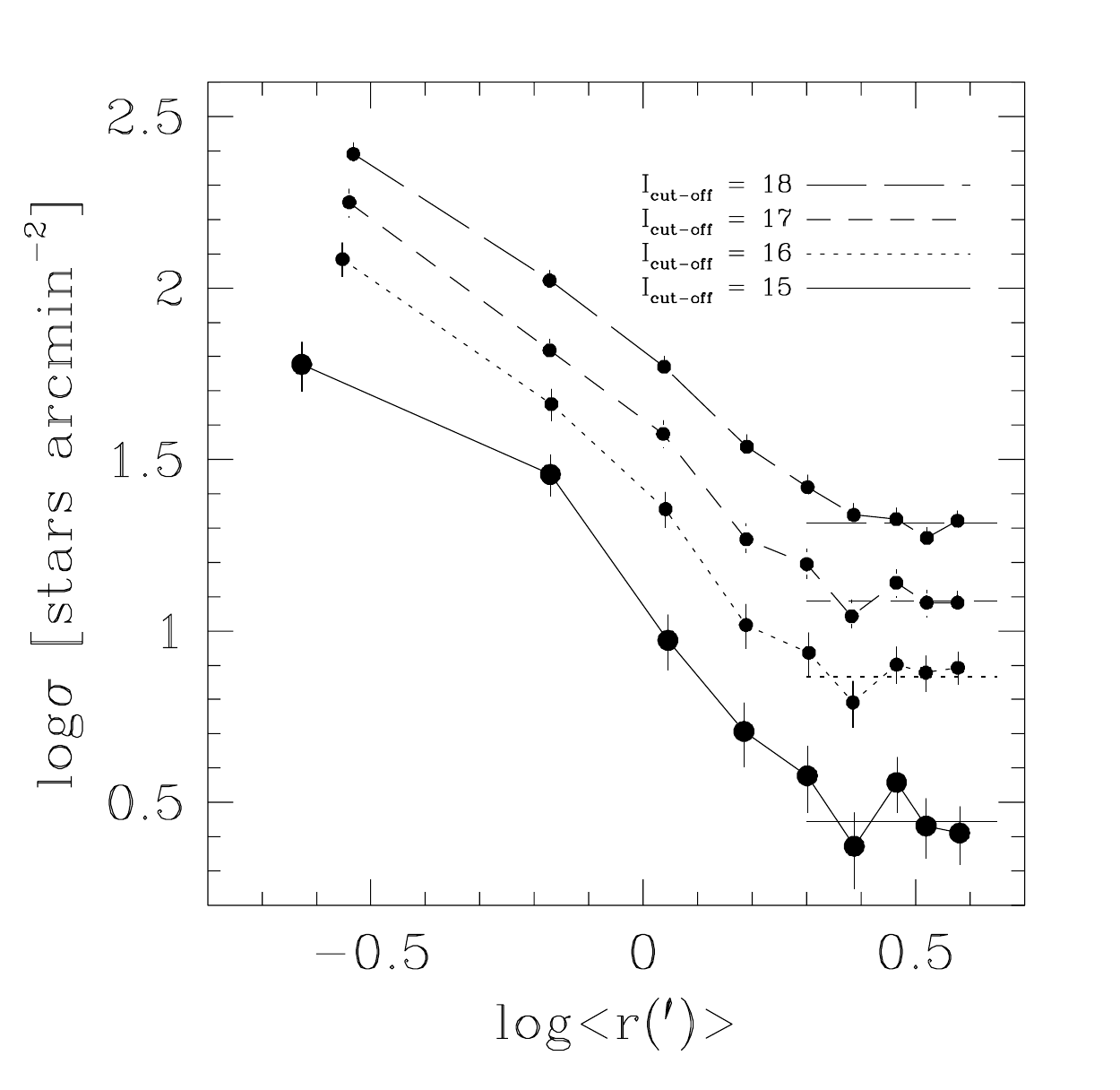}
\caption{The surface density profiles of Westerlund 1 with several 
cut-off $I$ magnitude. The horizontal line represents the average surface density 
of the field region within a given cut-off magnitude.
As the completeness of photometry for $I <$ 18 mag is over 80\% even in 
the cluster center, the incompleteness of data is not corrected.}
\label{fig13}
\end{figure}

The slope of the IMF in Fig.~\ref{fig11} seems to be divided into three parts;
$\log m \leq 1.0$, $1.0 < \log m \leq 1.4$, and $\log m > 1.4$. Given 
the result of \citet{GBSH11}, the slope in the mass range $\log m < 1.4$ 
may be similar to the Salpeter/Kroupa IMF or a slightly steeper IMF. 
For $1.0 < \log m \leq 1.4$, the IMF shows a plateau, while the slope of 
the IMF for the most massive stars ($\log m > 1.4$) is very steep, as seen 
in other starburst clusters, e.g. \objectname{the Arches cluster} 
\citep{KFKN07,FKMSRM99} and \objectname{NGC 3603} \citep{SB04}. However,  
the existence of a plateau is not certain. We can expect that the IMF 
for low-masses ($1.0 \lesssim \log m \lesssim 1.2$) may be lower than 
the present value in Fig.~\ref{fig11} 
if the IMF is compared to that of \objectname{NGC 3603} or that derived 
from NIR photometry. In general, the distribution of interstellar material 
within the Galactic disk is complex. Moreover, Westerlund 1 resides in 
the Scutum-Centaurus arm beyond the Sagittarius arm, and thus a non-uniform 
surface density of field stars is to be expected due to the superposition 
of two different spiral arms in the line of sight. Although the assumption 
that field stars are uniformly distributed in a given field of view is 
invariably assumed in order to select cluster members without independant 
membership criteria, a number of photometric studies of open clusters actually 
suffer from this statistical problem, to a greater or lesser degree. 
If the surface density of stars outside of the cluster region was 
underestimated, the resulting IMF would be steeper than the true IMF. 
In addition, reddening corrections and increasing photometric errors for 
faint stars could also contribute to the uncertainty in mass due to high 
levels of extinction, and the result may influence the shape of the IMF. 
On the other hand, the IMF obtained from NIR photometry for stars with 
masses below 10 $M_{\sun}$ may be much more reliable than that from the 
optical photometry. Given that the shape of the IMF ($\log m \sim 1$) is 
similar to that of \objectname{NGC 3603}, if the presence of the plateau 
in the IMF is real, other interpretations could be the presence of changes 
of slope in the IMF and/or a dependence on the calibration schemes.
  
There is practically no discussion in the literature on the presence of 
changes of slope in the IMF in high mass regimes, because in most young open 
clusters the number of massive stars ($m \geq 10M_{\sun}$) is very small. 
However, the number of massive stars in starburst clusters is not small, and 
so the result from such clusters is statistically meaningful.

\citet{DKS07} attempted to reproduce the IMF of \objectname{the Arches cluster}
from semi-analytical modelling. According to their work, the shape of 
the stellar IMF could be a result of the coalescence of pre-stellar cores 
(PSCs) at an early stage of cluster formation. The number of intermediate-mass 
PSCs rapidly decreases as they contract or merge with more massive PSCs. 
As a result, the stellar IMF of \objectname{the Arches cluster} reveals 
a plateau in the mass range 3--10$M_{\sun}$. In the case of Westerlund 1, 
despite its similar shape IMF, the plateau appears in the mass range 10 --
25 $M_{\sun}$. If the plateau in the IMF of Westerlund 1 is related to 
the depletion of intermediate-mass PSCs during the cluster formation stage 
as suggested by \citet{DKS07}, a different process may have operated in 
Westerlund 1. If we compare the age of three starburst clusters 
(\objectname{the Arches cluster}, \objectname{NGC 3603}, and Westerlund 1), 
the time evolution of the IMF can be excluded because the IMF of 
\objectname{NGC 3603} is very similar to that of Westerlund 1, but the age of 
\objectname{NGC 3603} is comparable to that of \objectname{the Arches cluster}.
On the other hand, \objectname{NGC 3603} and Westerlund 1 are located in 
a spiral arm, while \objectname{the Arches cluster} is at the Galactic center. 
If merging among PSCs effectively occurs near the Galactic 
center due to differences in the star formation environment, such as strong 
tidal forces, the plateau could have occurred at a relatively low-mass range.

Several tables providing the relation between temperature and bolometric 
correction have been published. Unfortunately, a reliable calibration covering 
the complete spectral ranges is very rare. Due to the lack of many 
representative O stars and the use of inadequate LTE stellar atmospheric 
models for early O type stars, especially for supergiants, prior to 2000, 
only limited information was available at the time. As the number of observed 
massive objects has increased and new atmospheric models taking into account 
non-LTE have emerged, a new and more reliable temperature scale and set of 
bolometric corrections for O3--O9.5 stars have been provided by \citet{MSH05}. 
The temperature scale of \citet{MSH05} is systematically lower than that of 
\citet{f77}, \citet{B81} or \citet{C97} for O type stars. Fortunately, since 
Flower's temperature scale is very similar to that of \citet{MSH05} for 
late O type stars, a small discrepancy does not seriously affect our 
results. However, it is clear that discrepancies in effective temperature 
can be a factor resulting in systematic differences in bolometric correction, 
thereby producing different individual stellar masses. And therefore, 
the changes of the slope in the IMF of Westerlund 1 and \objectname{NGC 3603} 
may be caused by the adopted temperature scale and bolometric correction 
calibration. In addition, in the calculation of the IMF of \objectname{NGC 3603}, 
\citet{SB04} used the mass-luminosity relation from the stellar evolution 
models of \citet{SSMM92}, if the changes of the slope is not a real feature, 
it may also be caused by the mass-luminosity relation of massive stars used by 
the Geneva group.

\subsection{The Abnormal Color of W43c}
The ($B-V$) color of the O type supergiant W43c is very abnormal 
($B-V \sim 2.0$; see Fig.~\ref{fig4}), but its ($V-I$) is fairly 
normal in the ($V,V-I$) CMD. When we adopt the 
mean reddening of $<E(B-V)>=4.19$ mag, the intrinsic color 
of the star is -2.18 mag. Given the intrinsic color 
$(B-V)_0\sim-0.33$ mag of the hottest stars, a $B-V$ of -2.18 
cannot be reached for a normal star. To identify 
the source of discrepancy, we have checked the optical images and 
photometric data.

The abnormal $B-V$ color is not caused by photometric error 
because the star is bright. Our $B-V$ of the star is 
consistent with that of \citet{NCR10} (See Table $1$ of their paper). 
The residual $V$ band image after subtracting all the measured stars 
shows that W43c is an optical double star, while the $B$ band images 
show no sign of an optical double. Hence, we can attribute the 
cause of abnormal ($B-V$) color to the superposition of a foreground 
star and a member (O9 Ib) of Westerlund 1. To confirm this we made a simple 
calculation by assuming that the purported stars have the same flux in the 
$V$ band because the two components in the residual image are very similar. 
The magnitude of W43c in the $V$ band was resolved 
into two using Pogson's formula. We also assume that the field star 
has $B-V=1.3$ mag and $V-I=2.0$ mag, whereas the member of Westerlund 1 has 
$B-V=4.0$ mag and $V-I=5.0$ mag. The deconvolution of the $V$ magnitude 
yields $B$ and $I$ magnitudes for each star, finally we calculated 
the composite magnitude in $B$ and $I$ of the system. As a result, the expected 
magnitudes are in good agreement with the observed ones, 
i.e. $\Delta B = -0.03$ mag and $\Delta I = -0.02$ mag, where 
$\Delta$ is the expected magnitude minus the observed one. Consquently, 
the abnormal $B-V$ color of W43c can be interpreted as the result of 
the superpostion of a less-reddened foreground star and a highly 
reddened O type star in Westerlund 1.

\section{Summary}
Starburst clusters are important objects to understand the 
properties, evolution, and formation of massive stars as well as 
the star formation in the early universe. Among several starburst 
clusters in the Galaxy, Westerlund 1 is the closest and the most massive 
cluster. We carried out $BVI_C$ and $JK_{S}$ photometry of Westerlund 1. 

Using the spectral type-color relation we obtained 
the color excess and total extinction in $V$, i.e. $<E(B-V)>=4.19
\pm0.23$ mag and $<A_V> = 13.0 \pm 0.7$ mag, respectively. The interstellar 
extinction in the NIR was also derived using the reddening law of 
\citet{KFL06} and the empirical fiducial line (EFL) constructed in the $[M_{K_{S}},
(J-K_{S})_0]$ plane, i.e. $<E(J-K_{S})>=1.70\pm0.21$ mag and 
$A_{K_{S}}=1.12\pm0.14$ mag. Since the normal ZAMS fitting method was nearly impossible 
to apply to Westerlund 1 due to severe extinction, we fitted the EFL to 
the turn-on point from the PMS to the MS stars, thereby obtaining a distance of 
3.8 kpc. 

We constructed HR diagrams using the conversion table of \citet{f77}. By 
superimposing the new stellar evolution models with rotation of \citet{rot12} on the HR diagrams, 
an age of $5.0\pm1.0$ Myr was obtained. The result is in good agreement 
with previous studies. However, the presence of RSGs could imply a large 
spread in the age of Westerlund 1. Hence, several plausible star formation 
histories were suggested. 

The masses of evolved stars with spectral types were determined using the 
mass-luminosity relation provided by the stellar evolution models with 
rotation \citep{rot12}. We also determined the mass of individual stars 
without spectral types from the relation between absolute magnitude 
and initial mass. Finally, the IMF of Westerlund 1 was obtained after 
statistically subtracting the contribution of field interlopers to 
the mass function. The slope of the IMF was found to be as shallow as that of 
\objectname{NGC 3603} center, when we used the complementary IMF 
derived from NIR photometry. We discussed possible changes of the slope 
in the IMF in different mass regimes in Westerlund 1. 

The integration of the IMFs in the mass range ($85 M_{\sun} > m > 0.08 M_{\sun}$) 
yielded a total masses of $7.3 \times 10^4 M_{\sun}$ and $7.8 \times 10^4 M_{\sun}$, 
respectively, where we assumed that the shape of the IMF in low-mass regime may be similar 
to that of \objectname{NGC 2264} or that of Kroupa. Also, a lower limit of the total mass 
($5.1 \times 10^4 M_{\sun}$) was derived from the radial variation of the IMF. 
If we assume this mass as the total mass, a half-mass radius is 1.1 pc at 3.8 kpc. 
We confirm that Westerlund 1 is the most massive open cluster in the Galaxy.

The mass segregation of Westerlund 1 was also investigated and we found that 
the slope of the IMF varied systematically from the center to 
$r \sim 1\farcm7$, but the pattern of variation changed at $r \simeq 1\farcm7$. 
Some discussion was made concerning this issue.

\acknowledgments
This work was partly supported by a National Research Foundation of Korean (NRF) 
grant funded by the Korea Government (MEST) (Grant No.20120005318) and 
partly supported by the Korea Astronomy and Space Science Institute (KASI) (Grant No. 2011-9-300-02).

{\it Facilities: CTIO 4m, AAT 3.9m}

\appendix

\section{The red leak in the SDSS $u$ filter}
\begin{figure}
\epsscale{.80}
\plotone{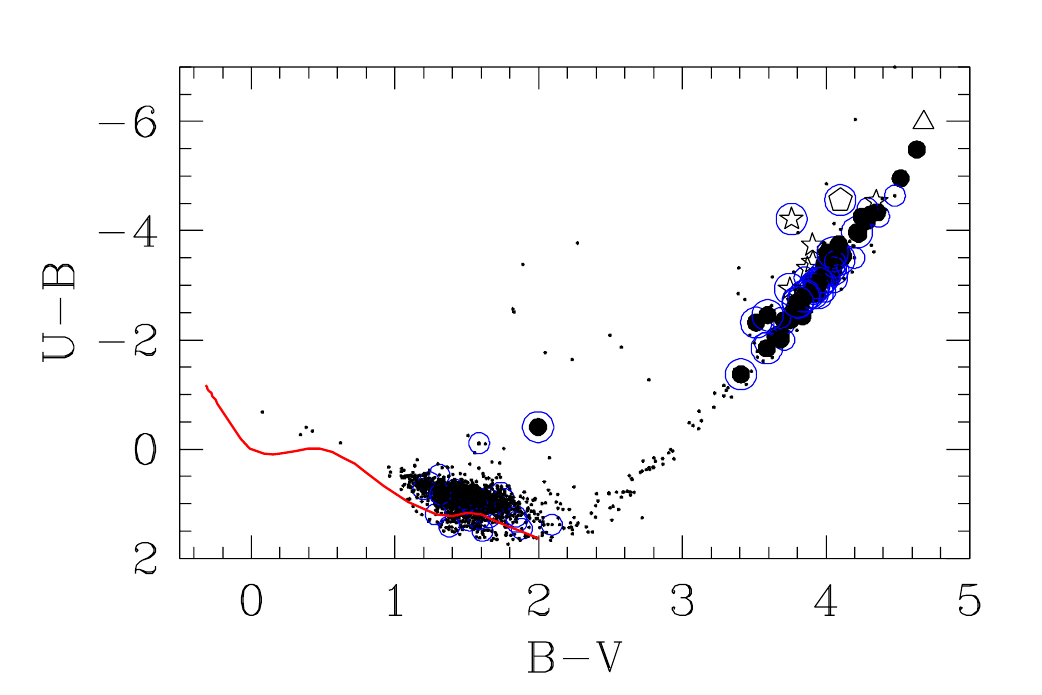}
\caption{The effect of the red leak in the SDSS $u$ filter used with the MOSAIC II CCD 
attached to the Blanco 4m telescope at CTIO. 
The solid line represents the intrinsic color relation for MS stars. 
There are two significant groups of stars in this diagram. The bluer group (in $B-V$) may be a
less reddened foreground population, while the 
redder group (in $B-V$) are suspected members of Westerlund 1. For those stars with $B-V > 2.0$, 
the $U-B$ color indices become increasingly bluer due to the red 
leak of the SDSS $u$ filter. } 
\label{fig14}
\end{figure}

Photometry in the $U$ band is of crucial important in reddening determination. 
Westerlund 1 is very reddened, but we had expected to detect the $U$ signal for a few of the brightest stars in Westerlund 1 with 
an exposure time of 1 hour in the $u$ band. Contrary to 
our expectations, there were many bright stars suspected as members of 
Westerlund $1$ in the images even in a 10-min exposed image. Interestingly, extremely red stars revealed a double-peak profile in the surface plot. The difference in brightness between the two peaks was not as great as 0.1 mag, thus photometric data in the $u$ band were seen to be 
dominated by the red light leak, either resolved or not. Unfortunately, reliable extremely
red standard stars for correcting this effect were absent. We confirmed the effect of the 
red leak on the ($U-B,B-V$) diagram (Fig.~\ref{fig14}). The suspected members with the redder 
color strongly suffered from the red leak in the $U-B$ color. \citet{B05} mentioned that 
the red leak could be suppressed by adding a cut-off filter or by using a 
leak-free coated filter. In the work, the author also presented a procedure for correcting 
for the red leak. Unfortunately, due to the heavy reddening of Westerlund 1, the resultant
$U-B$ data would be very uncertain, even were the red leak corrected by 
the procedure. 

\section{The empirical fiducial line}
The ZAMS relation is the most powerful tool in determining the distance 
to open clusters. Since the 
relation is based on the intrinsic properties of MS stars, that gives 
very accurate distance if the reddening toward a cluster 
is well determined. Unfortunately, the heavy extinction of Westerlund 1 
does not permit us to determine the accurate distance using the ZAMS, 
or theoretical isochrones, because our detection limit ($V \sim 24$ mag) 
is not faint enough to reach the gradual turning of the ZAMS, equivalent to A--F type stars. For this reason, we had to perform NIR photometry.

\begin{figure}
\epsscale{.80}
\plotone{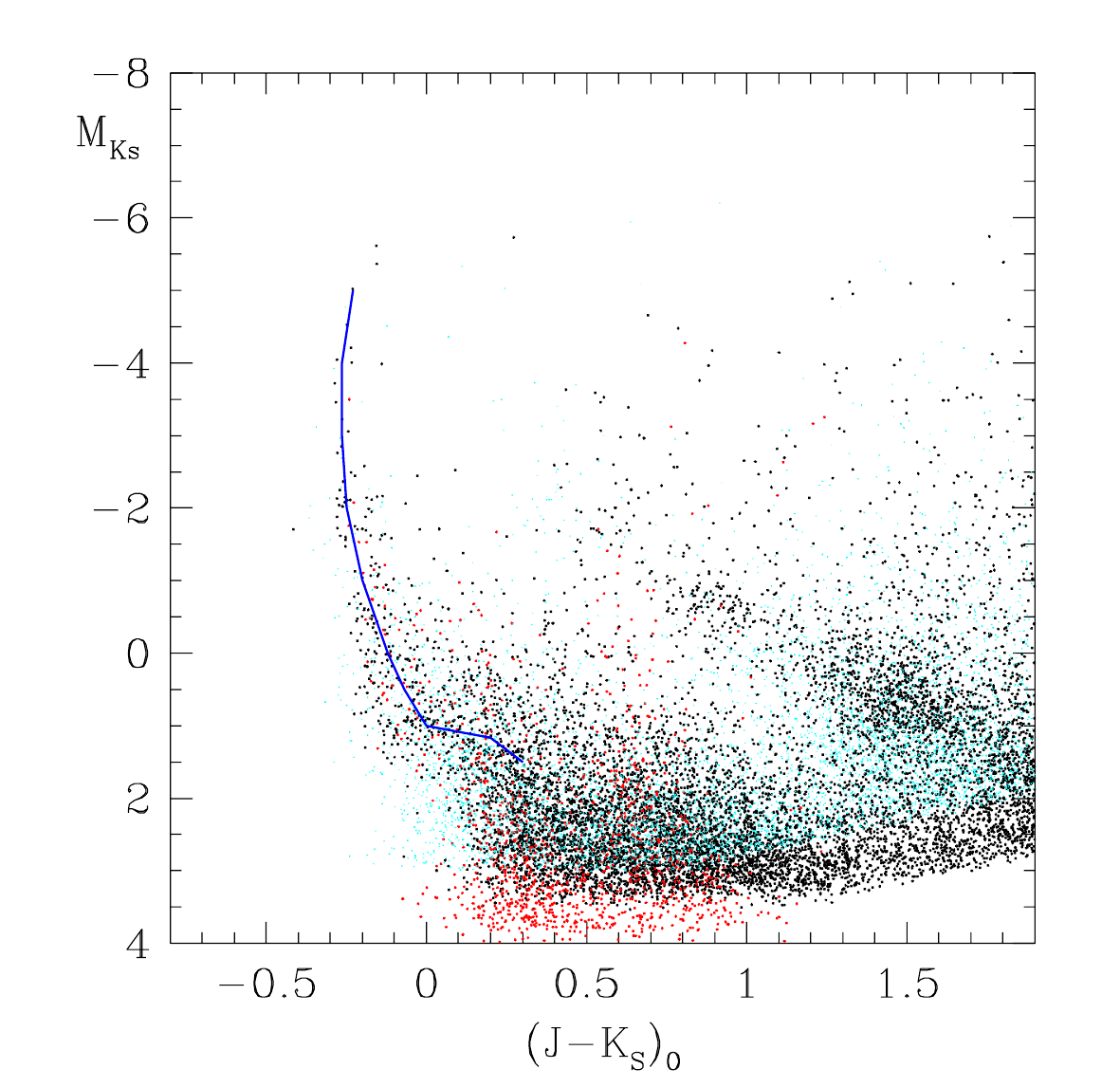}
\caption{The $M_{K_S}$ versus $(J-K_S)_0$ color-magnitude diagram of three open 
clusters. Red, black, and cyan represent 
the stars in NGC 2362, NGC 6231, and NGC 6823, respectively. We set an empirical 
fiducial line (solid line) based on the 
photometric data. See Appendix B for details.}
\label{fig15}
\end{figure}

What will be used here is similar to the isochrone placement as shown 
in \citet{BCSWNG08} and \citet{GBSH11}, but a more empirical approach as for the ZAMS 
relation. Firstly, we suppose the age of Westerlund 1 to be 3.0--6.0 Myr from previous 
studies. Open clusters with a similar age were searched for. \objectname{NGC 2362} 
(3.0--6.0 Myr -- \citet{M01}), \objectname{NGC 6231} (2.5-5.0 Myr -- \citet{SB93,SBL98,BVF99}), 
and \objectname{NGC 6823} (2.0-7.0 Myr -- \citet{BBD08}) 
were selected for constructing the empirical relation. Subsequently, $UBV$ CCD 
photometric data of each of the clusters were obtained from WEBDA 
(http://www.univie.ac.at/webda/), and we adopted the distance and reddening from \citet{D06} 
for \objectname{NGC 2362}, \citet{SBL98} for \objectname{NGC 6231}, 
and \citet{G92} for \objectname{NGC 6823}. NIR data for these clusters 
were obtained from the 2MASS catalogue \citep{2mass}. To place NIR photometric 
data of stars belonging to the cluster in the [$K_{S},(J-K_{S})_0$] plane 
we needed to determine the reddening in the NIR region. Using the intrinsic 
color relations, i.e. $(B-V)_0$ versus $(V-\lambda)_0$, $E(V-J)$ and 
$E(V-K_{S})$ were determined, thereby $E(J-K_S)$ and $A_{K_{S}}$ were obtained by adopting 
the $A_V$ values from the literature. We presented the results in Fig.~\ref{fig15}. 
In the case of \objectname{NGC 2362}, the reddening is so small that 
the total extinction $A_{K_{S}}$ would be strongly affected by statistical 
fluctuations, so we assumed $A_{K_{S}}=0.0$ for \objectname{NGC 2362}. As all NIR data of the clusters 
were corrected for reddening and distance, the CMDs could be superimposed 
in the photometric plane on an absolute scale. Finally, we drew the 
best fitting line which traced the middle of the MS and PMS sequences and presented it 
in Fig.~\ref{fig15}. We call this the empirical fiducial line (EFL).

\clearpage

\clearpage

\clearpage

\clearpage

\clearpage

\clearpage

\clearpage

\clearpage

\clearpage

\clearpage

\clearpage

\clearpage

\clearpage

\clearpage

\clearpage

\clearpage









\begin{deluxetable}{cccc}
\tabletypesize{\scriptsize}
\tablewidth{0pt}
\tablecaption{Atmospheric Extinction Coefficients. \label{tab1}}
\tablehead{
\colhead{filter} & \colhead{$k_1$ (2009 Mar. 28)} & \colhead{$k_1$ (2009 Mar. 29)} & \colhead{$k_2$} }
\startdata
$B$ & $0.244\pm0.010$ &\nodata          & $0.026\pm0.001$ \\
$V$ & $0.141\pm0.006$ &$0.134\pm0.014$ & \nodata \\
$I_C$ & $0.053\pm0.011$ &$0.075\pm0.018$ & \nodata \\
\enddata
\end{deluxetable}
\clearpage

\begin{deluxetable}{ccccccc}
\tabletypesize{\scriptsize}
\tablewidth{0pt}
\tablecaption{Coefficients for Mosaic II data \label{tab2}}
\tablehead{\colhead{Filter} & \colhead{Chip} & \colhead{$\eta_{1}$} &\colhead{$\eta_{2}$\tablenotemark{1}} 
& \colhead{$\beta$} &\colhead{$\gamma$} & \colhead{$\zeta$ (2009 Mar. 28)}}
\startdata
       &  1   & $ 0.046\pm0.008$ &  \nodata   & $ -.012\pm0.003$ & $ -.018\pm0.002$ & $25.481\pm0.009$ \\
       &  2   & $ 0.041\pm0.006$ &  \nodata   & $ -.005\pm0.003$ & $ -.013\pm0.001$ & $25.444\pm0.010$ \\
       &  3   & $ 0.034\pm0.005$ &  \nodata   & $ 0.004\pm0.002$ & $ -.011\pm0.001$ & $25.432\pm0.007$ \\
 $B$   &  4   & $ 0.116\pm0.003$ &  \nodata   & $ 0.003\pm0.003$ & $ -.013\pm0.002$ & $25.404\pm0.009$ \\
       &  5   & $ 0.058\pm0.005$ &  \nodata   & $ -.010\pm0.003$ & $ 0.001\pm0.002$ & $25.412\pm0.010$ \\
       &  6   & $ 0.063\pm0.004$ &  \nodata   & $ -.006\pm0.003$ & $ 0.003\pm0.001$ & $25.383\pm0.009$ \\
       &  7   & $ 0.057\pm0.003$ &  \nodata   & $ 0.010\pm0.002$ & $ 0.007\pm0.001$ & $25.361\pm0.009$ \\
       &  8   & $ 0.087\pm0.003$ &  \nodata   & $ 0.007\pm0.003$ & $ -.001\pm0.002$ & $25.376\pm0.010$ \\
\hline
       &  1   & $ -.025\pm0.002$ &  $0.000$   & $ 0.003\pm0.001$ & $ -.016\pm0.001$ & $25.695\pm0.009$ \\
       &  2   & $ -.018\pm0.002$ &  $0.000$   & $ -.007\pm0.001$ & $ -.015\pm0.001$ & $25.692\pm0.009$ \\
       &  3   & $ -.020\pm0.002$ &  $0.000$   & $ 0.001\pm0.001$ & $ -.016\pm0.001$ & $25.676\pm0.007$ \\
 $V$   &  4   & $ -.023\pm0.002$ &  $0.000$   & $ 0.007\pm0.001$ & $ -.017\pm0.001$ & $25.688\pm0.009$ \\
$(B-V)$&  5   & $ -.036\pm0.002$ &  $0.000$   & $ -.001\pm0.001$ & $ 0.005\pm0.001$ & $25.647\pm0.010$ \\
       &  6   & $ -.026\pm0.003$ &  $0.000$   & $ 0.003\pm0.001$ & $ 0.005\pm0.001$ & $25.627\pm0.010$ \\
       &  7   & $ -.029\pm0.001$ &  $0.000$   & $ 0.008\pm0.001$ & $ 0.005\pm0.001$ & $25.622\pm0.009$ \\
       &  8   & $ -.030\pm0.002$ &  $0.000$   & $ 0.013\pm0.001$ & $ 0.007\pm0.001$ & $25.627\pm0.009$ \\
\hline
       &  1   & $ -.023\pm0.003$ &  $0.000$   & $ 0.003\pm0.001$ & $ -.016\pm0.001$ & $25.697\pm0.008$ \\
       &  2   & $ -.014\pm0.003$ &  $0.000$   & $ -.007\pm0.001$ & $ -.015\pm0.001$ & $25.690\pm0.008$ \\
       &  3   & $ -.017\pm0.003$ &  $0.000$   & $ 0.001\pm0.001$ & $ -.016\pm0.001$ & $25.676\pm0.008$ \\
 $V$   &  4   & $ -.018\pm0.004$ &  $0.000$   & $ 0.007\pm0.001$ & $ -.017\pm0.001$ & $25.685\pm0.009$ \\
$(V-I)$&  5   & $ -.031\pm0.003$ &  $0.000$   & $ -.001\pm0.001$ & $ 0.005\pm0.001$ & $25.646\pm0.010$ \\
       &  6   & $ -.020\pm0.003$ &  $0.000$   & $ 0.003\pm0.001$ & $ 0.005\pm0.001$ & $25.625\pm0.010$ \\
       &  7   & $ -.024\pm0.002$ &  $0.000$   & $ 0.008\pm0.001$ & $ 0.005\pm0.001$ & $25.622\pm0.009$ \\
       &  8   & $ -.028\pm0.003$ &  $0.000$   & $ 0.013\pm0.001$ & $ 0.007\pm0.001$ & $25.630\pm0.009$ \\
\hline
       &  1   & $ -.016\pm0.002$ &  \nodata   & $ -.003\pm0.002$ & $ -.018\pm0.001$ & $25.131\pm0.010$ \\
       &  2   & $ -.025\pm0.001$ &  \nodata   & $ -.013\pm0.001$ & $ -.016\pm0.001$ & $25.109\pm0.009$ \\
       &  3   & $ -.012\pm0.003$ &  \nodata   & $ 0.001\pm0.001$ & $ -.022\pm0.001$ & $25.111\pm0.010$ \\
 $I_C$ &  4   & $ -.014\pm0.002$ &  \nodata   & $ 0.001\pm0.001$ & $ -.015\pm0.001$ & $25.100\pm0.010$ \\
       &  5   & $ -.020\pm0.002$ &  \nodata   & $ -.008\pm0.001$ & $ 0.006\pm0.001$ & $25.081\pm0.010$ \\
       &  6   & $ -.021\pm0.002$ &  \nodata   & $ -.010\pm0.001$ & $ 0.004\pm0.001$ & $25.047\pm0.010$ \\
       &  7   & $ -.010\pm0.001$ &  \nodata   & $ 0.014\pm0.001$ & $ 0.006\pm0.001$ & $25.011\pm0.010$ \\
       &  8   & $ -.019\pm0.002$ &  \nodata   & $ 0.008\pm0.001$ & $ 0.003\pm0.001$ & $25.050\pm0.010$ \\
\enddata
\tablenotetext{1}{$\eta_{2}$ for $V$ represents the transformation coefficient for $B-V > 1.5$ and $V-I > 1.5$. }
\end{deluxetable}
\clearpage

\begin {deluxetable}{lcccccc}
\rotate
\tabletypesize{\scriptsize}
\tablewidth{0pt}
\tablecaption{Comparison of Photometry\label{tab3}}
\tablehead{
Paper          &
$\Delta V$\tablenotemark{1}     & N(m)\tablenotemark{2} &
$\Delta$(\bv)\tablenotemark{1}  & N(m)\tablenotemark{2} &
$\Delta (V-I)$\tablenotemark{1} & N(m)\tablenotemark{2} }
\startdata
\citet{PBC98}  & $-1.06\pm0.13$ & 22 (2) &                &        & $0.24\pm0.13$ & 11 (0) \\
\citet{CNCG05} & $-0.42\pm0.10$ & 48 (3) & $ 0.12\pm0.15$ & 45 (2) & $0.21\pm0.08$ & 33 (2) \\
\citet{B07}    & $-0.05\pm0.06$ & 19 (1) & $-0.02\pm0.01$ &  2 (0) & $0.61\pm0.10$ &  3 (0) \\
\enddata
\tablenotetext{1}{Others - This}
\tablenotetext{2}{N and m represents number of compared stars and excluded stars}
\end{deluxetable}
\clearpage

\begin{deluxetable}{rcccccccccccccccc}
\rotate
\tabletypesize{\scriptsize}
\tablewidth{0pt}
\tablecaption{Photometric Data\label{tab4}}
\tablehead{
\colhead{ID\tablenotemark{a}} & \colhead{$\alpha_{J2000}$}  & \colhead{$\delta_{J2000}$} &
\colhead{$V$} & \colhead{$I$} & \colhead{$V-I$} & \colhead{$B-V$} & \colhead{$U-B$} & 
\colhead{$\epsilon_V$} & \colhead{$\epsilon_I$} & \colhead{$\epsilon_{V-I}$} & 
\colhead{$\epsilon_{B-V}$} & \colhead{$\epsilon_{U-B}$} & \colhead{N$_{obs}$} & 
\colhead{2MASS } & \colhead{remark}\tablenotemark{b} & \colhead{Sp}\tablenotemark{c}}
\startdata
14846&16 47 02.14&-45 51 43.9& 20.941& 16.078&  4.848&  3.436& -2.737&  0.022&  0.021&  0.030&  0.109&  0.127&3 1 1 1 1&       -        &  &        \\
14847&16 47 02.15&-45 49 09.8& 23.522& 17.699&  5.823&\nodata&\nodata&  0.002&  0.004&  0.005&\nodata&\nodata&3 4 3 0 0&16470214-4549097&  &        \\
14848&16 47 02.15&-45 49 33.2& 23.223& 19.926&  3.297&\nodata&\nodata&  0.004&  0.013&  0.013&\nodata&\nodata&3 4 3 0 0&       -        &  &        \\
14849&16 47 02.15&-45 48 46.3& 23.760& 20.594&  3.165&\nodata&\nodata&  0.009&  0.012&  0.015&\nodata&\nodata&3 3 3 0 0&       -        &  &        \\
14850&16 47 02.15&-45 51 12.6& 19.125& 13.781&  5.246&  4.093& -3.472&  0.002&  0.009&  0.009&  0.003&  0.005&3 1 1 3 2&16470215-4551126& X& O9.0Iab\\
14851&16 47 02.15&-45 51 40.7& 23.200&\nodata&\nodata&\nodata&\nodata&  0.021&\nodata&\nodata&\nodata&\nodata&3 0 0 0 0&       -        &  &        \\
14852&16 47 02.15&-45 52 11.6& 22.094& 19.502&  2.592&  2.065&\nodata&  0.013&  0.008&  0.015&  0.014&\nodata&3 4 3 3 0&       -        &  &        \\
14853&16 47 02.15&-45 50 40.8&\nodata& 21.156&\nodata&\nodata&\nodata&\nodata&  0.022&\nodata&\nodata&\nodata&0 2 0 0 0&       -        &  &        \\
14854&16 47 02.15&-45 46 51.6&\nodata& 21.711&\nodata&\nodata&\nodata&\nodata&  0.055&\nodata&\nodata&\nodata&0 3 0 0 0&       -        &  &        \\
14855&16 47 02.16&-45 49 53.1& 23.795& 20.786&  3.010&\nodata&\nodata&  0.022&  0.014&  0.026&\nodata&\nodata&3 3 3 0 0&       -        &  &        \\
14856&16 47 02.17&-45 51 54.4&\nodata& 20.884&\nodata&\nodata&\nodata&\nodata&  0.041&\nodata&\nodata&\nodata&0 3 0 0 0&       -        &  &        \\
14857&16 47 02.17&-45 50 21.5& 23.299& 18.423&  4.876&\nodata&\nodata&  0.019&  0.033&  0.038&\nodata&\nodata&3 4 3 0 0&       -        &  &        \\
14858&16 47 02.17&-45 47 21.3& 17.277& 15.644&  1.624&  1.293&  0.703&  0.006&  0.008&  0.010&  0.006&  0.003&3 1 1 3 2&16470217-4547211&  &        \\
14859&16 47 02.17&-45 51 07.6& 22.693& 19.586&  3.107&\nodata&\nodata&  0.002&  0.092&  0.092&\nodata&\nodata&3 1 1 0 0&       -        &  &        \\
14860&16 47 02.17&-45 55 03.1& 18.760& 16.975&  1.774&  1.339&  0.607&  0.006&  0.009&  0.011&  0.007&  0.003&3 4 3 3 2&       -        &  &        \\
14861&16 47 02.17&-45 49 44.6& 18.646& 16.817&  1.820&  1.448&  0.830&  0.001&  0.003&  0.003&  0.006&  0.006&3 4 3 3 2&       -        &  &        \\
14862&16 47 02.18&-45 51 30.5& 22.649& 17.425&  5.224&\nodata&\nodata&  0.013&  0.014&  0.019&\nodata&\nodata&3 1 1 0 0&       -        &  &        \\
\enddata
\tablecomments{
Table \ref{tab4} is published in its entirety in the
electronic edition with the recalibrated photometric data from \citet{CNCG05,NCR10} and \citet{PBC98}.  A portion is shown here for guidance regarding its form and content.}
\tablenotetext{a}{The negative numbered ID represents the calibrated data 
from \citet{CNCG05,NCR10,PBC98}}
\tablenotetext{b}{X: X-ray emission stars, x: X-ray emission candidate}
\tablenotetext{c}{Spectral type - \citet{CN02,CN04,CNCG05,CHCNV06,MT07,NCR10,CRN10,DCNJC10}}
\end{deluxetable}
\clearpage



\end{document}